\documentclass[a4paper,11pt]{article}
\pdfoutput=1 
\usepackage{jcappub} 

\usepackage[T1]{fontenc}
\usepackage{comment}
\usepackage{caption}
\usepackage{subcaption}
\usepackage{aas_macros}
\usepackage[dvipsnames]{xcolor}
\usepackage[capitalise]{cleveref}

\newcommand{\be}{\begin{equation}}
\newcommand{\ee}{\end{equation}}
\newcommand{\bea}{\begin{eqnarray}}
\newcommand{\eea}{\end{eqnarray}}
\newcommand{\bfk}{\boldsymbol{k}}
\newcommand{\bfr}{\boldsymbol{r}}
\newcommand{\bfs}{\boldsymbol{s}}
\newcommand{\bfx}{\boldsymbol{x}}

\newcommand{\zhat}{\hat{\boldsymbol{z}}}
\newcommand{\bfw}{\boldsymbol{w}}
\newcommand{\bfW}{\boldsymbol{W}}

\newcommand{\pl}{P_\mathrm{lin}}
\newcommand{\plir}{P_\mathrm{lin,IR}}

\newcommand{\rmr}{\mathrm{r}}
\newcommand{\rms}{\mathrm{s}}
\newcommand{\xir}{\xi}
\newcommand{\xirud}{\xi_{12}}
\newcommand{\xirdt}{\xi_{23}}
\newcommand{\xirtu}{\xi_{31}}
\newcommand{\xis}{\xi_\rms}
\newcommand{\zetar}{\zeta}
\newcommand{\zetas}{\zeta_\rms}
\newcommand{\triangler}{\triangle_{\rmr}}
\newcommand{\triangles}{\triangle_{\rms}}
\newcommand{\lo}{\stackrel{\mathrm{LO}}{\simeq}}

\newcommand{\sst}{\stackrel{\mathrm{SSt}}{=}}

\title{\boldmath The streaming model for the three-point correlation function and its connection to standard perturbation theory}

\author[a]{A. Pugno,}
\author[a]{A. Eggemeier,}
\author[a,b,c,d]{C. Porciani}
\author[e]{and J. Kuruvilla}

\affiliation[a]{Argelander Institut f\"ur Astronomie, Auf dem H\"ugel 71, Bonn, Germany}
\affiliation[b]{SISSA, International School for Advanced Studies, Via Bonomea 265, 34136 Trieste, TS, Italy}
\affiliation[c]{ Dipartimento di Fisica – Sezione di Astronomia, Università di Trieste, Via Tiepolo 11, 34131 Trieste, Italy}
\affiliation[d]{IFPU, Institute for Fundamental Physics of the Universe, via Beirut 2, 34151 Trieste, Italy}
\affiliation[e]{Universit\'e Paris-Saclay, CNRS, Institut d’astrophysique spatiale, 91405, Orsay, France}
\emailAdd{apugno@uni-bonn.de}

\abstract{
Redshift-space distortions (RSDs) present a significant challenge in building models for the three-point correlation function (3PCF).
We compare two possible lines of attack: the streaming model and standard perturbation theory (SPT). The two approaches differ in their treatment of the non-linear mapping from real to redshift space: SPT expands this mapping perturbatively, while the streaming model retains its non-linear form but relies on simplifying assumptions about the probability density function (PDF) of line-of-sight velocity differences between pairs or triplets of tracers. 
To assess the quality of the predictions and the validity of the assumptions of these models, we measure the monopole of the matter 3PCF and the first two moments of the pair- and triplewise velocity PDF from a suite of N-body simulations. 
We also evaluate the large-scale limit of the streaming model and determine under which conditions it aligns to SPT. 
On scales larger than $10\,h^{-1}\,\mathrm{Mpc}$, we find that the streaming model for the 3PCF monopole is dominated by the first two velocity moments, making the exact shape of the PDF irrelevant. 
This model can match the accuracy of a Stage-IV galaxy survey, if the velocity moments are measured directly from the simulations. However, replacing the measurements with perturbative expressions to leading order generates large errors already on scales of  $60$ - $70\,h^{-1}\,\mathrm{Mpc}$. This is the primary drawback of the streaming model. 
On the other hand, the SPT model for the 3PCF cannot account for the significant velocity dispersion that is present at all scales, and consequently provides predictions with limited accuracy.
We demonstrate that this issue can be approximately addressed by isolating the large-scale limit of the dispersion, which leads to typical Fingers-of-God damping functions. Overall, the SPT model with a damping function provides the best compromise in terms of accuracy and computing time.

}

\begin{document}
\maketitle
\flushbottom

\section{Introduction}
\label{sec:intro}

The current generation of galaxy redshift surveys such as 
the Dark Energy Spectroscopic Instrument \cite{desi2016, desi2016b}, 
Euclid \cite{Mellier+24}, and the Roman Space Telescope \cite{Eifler2021}, will observe unprecedentedly large comoving volumes and probe the large-scale structure (LSS) of the Universe with a high number density of tracers,
thus allowing accurate and precise measurements of the (connected) three-point correlation function (3PCF) and its Fourier transform, the bispectrum.
In order to make the best use of these Stage-IV surveys and extract the full cosmological information they encode, it is pivotal to be able
to model the signal expected from these higher-order (with respect to the power spectrum) clustering summary statistics 
\cite{Fry1994, Frieman1994, Matarrese1997, Sefusatti2006, Karagiannis2018, Yankelevich2018,  Marin2011, Amendola2013, Gil-Marin2015, Chudaykin2019, Gualdi2020, Hahn2020, Oddo2020, Oddo2021, Veropalumbo2021, Ivanov2022, Philcox2022, Frizzo2023, Farina2024}.

For Gaussian initial conditions, three-point statistics of the matter distribution are intrinsically non-linear and can be studied perturbatively in Fourier \citep{Fry1984, Scoccimarro1997, matsubara2008, Rampf2012} and configuration space \citep{jing1996, Frieman1994, Barriga2002}. The same methodology can be adopted to account for galaxy biasing and redshift-space distortions (RSDs) \citep{Fry1994, Hivon1995, ScoCouFri9906, Matarrese1997, Verde1998, Nishimichi2007,  SleEis1503, HasRasTar1708}. More recently, these approaches have been generalised to the effective field theory (EFT) of LSS \citep{Angulo2015, Baldauf2015b, debelsunce-senatore2019, Ivanov2022}. 
On large scales, perturbative
predictions in Fourier space are in good agreement with N-body simulations \citep{Lazanu2016, Steele2021, Ivanov2022, Alkhanishvili2022} and this motivates applying Bayesian inference based on these models to actual observational data \citep[e.g.][]{Gil-Marin2015,  Ivanov2023, Damico2024}.

Alternative theoretical approaches have been proposed, either based on the so-called halo model \cite{Takada2003, Smith2008} and/or by providing fitting
formulae for the matter bispectrum measured in N-body simulations \citep{Scoccimarro2001, Gil-Marin2012, Gil-MArin2014, Takahashi2020}.

In general, recent research has focused principally on the bispectrum with only a few investigations addressing the 3PCF in real- \citep{Hoffmann2018, Guidi2023} and redshift-space \citep{SleEis1503, sugiyama2021}.
This study aims to fill the gap in the literature by addressing RSDs in the 3PCF. As a first step, we concentrate on the matter distribution, thus neglecting the problem of galaxy biasing.

Peebles \citep{peebles1980} introduced a non-perturbative method to account for RSDs in the 2PCF, which is now known as the ``streaming model''. The model is exact in the distant-observer approximation, i.e.
when the line of sight (LOS) to all galaxies in a survey can be approximated
with a single unit vector.
The basic idea is to use an integral transformation of the real-space correlation function (CF) that depends on the probability density function (PDF) of the relative LOS velocities of galaxy (or particle) pairs 
\citep[see also][]{fisher1994, scoccimarro2004, kuruvilla2018, cuesta2020}. The model, however, needs to be informed about the shape
of the velocity PDF and the value of its moments in order to be predictive.
Some authors managed to combine this approach with Lagrangian perturbation theory \citep{reid2011} and EFT \cite{vlah2016} to make accurate predictions of the 2PCF in redshift space.

Recently, it has been demonstrated from first principles that similar integral transformations
also hold true for CFs of any order \citep{kuruvilla2020} but they
are more naturally written in terms of full CFs (instead of their connected parts). The price to pay for expressing the $n$-point
streaming model in terms of connected functions is that correlation
of different orders mix; for instance, the expression for the 3PCF in redshift space contains both the 3PCF and the 2PCF in real space
as well as the PDFs of the differences between peculiar velocities
of particles forming pairs and triplets.

In this paper, we use a large suite of N-body simulations to determine
the regime of validity of the models for the 3PCF in redshift space
based on standard perturbation theory (SPT) and the streaming model. 
For the latter, we assume either Gaussian or Laplace velocity PDFs
but we also determine the range of scales at which results do not
depend on the precise form of the PDF.

The paper is organised as follows. In Section~\ref{sec: models},
we provide the necessary theoretical background. The N-body simulations and the estimators we use to extract
the 2 and 3PCF are presented in Section~\ref{sec: data and measurements}.
Here, we also show the measurements of the moments of the relative velocities within triplets of particles 
and compare them to the SPT predictions at leading order.
In Section~\ref{sec: results}, we contrast the measurement of the redshift-space 3PCF from 
the N-body simulations with the model predictions from SPT and the
streaming model.
A systematic and comprehensive derivation of the large-scale limit
of the streaming model is presented in Section~\ref{sec: lsl}.
This provides us with the tools to match this model to the SPT expression term by term and understand what exactly generates differences between
the two schemes. In addition, we use this calculation to 
motivate the phenomenological damping functions often used in the literature to correct the SPT predictions at small scales for
the presence of multi-stream flows.
Finally, in Section~\ref{sec: conclusion}, we draw our conclusions.
Five appendices present more technical material that helps achieving a full understanding of the main text.

\section{Models for the three-point correlation function in redshift space}
\label{sec: models}

\subsection{Two- and three-point correlation functions}
\label{sec: rsd, 3pcf}

Given the local matter density $\rho(\bfx)$ and its mean value $\bar{\rho} = \langle \rho(\bfx)\rangle$, we define the two-point correlation function (2PCF) as the product between the density contrast, $\delta(\bfx) = \rho(\bfx) / \bar{\rho}-1$, evaluated at two different locations and averaged over a statistical ensemble of realisations:
\begin{equation}
\xir(r)=\langle \delta(\bfx) \,\delta(\bfx+\bfr) \rangle\,.
\end{equation}
Assuming that ensemble-averaged statistics are invariant under translations and rotations implies that $\xi$ depends only on the magnitude of the separation vector $r$.
Similarly, we introduce the three-point correlation function (3PCF),

\begin{equation}
\label{eq:3pcf ensamble}
    \zetar (r_{12}, r_{23}, r_{31})= \langle \delta(\bfr_1)\, \delta(\bfr_2) \,\delta(\bfr_3) \rangle \,,
\end{equation}
which, under the same assumptions of translation- and rotation-invariance, is fully characterized by the magnitudes $r_{ij}$ of the pair separations $\bfr_{ij}=\bfr_i-\bfr_j$
(with $\bfr_{12} + \bfr_{23} + \bfr_{31} =0$).
Instead of the three sidelengths of the triangular configuration, the 3PCF may also be parameterised in terms of only two sides and the cosine of the angle between them.

For our applications, it is convenient to consider the Fourier transform of the density contrast 
\begin{equation}
    \tilde{\delta}(\bfk)=\int \delta(\bfx)\,e^{-i\bfk\cdot\bfx}\,\mathrm{d}^3x\;,
\end{equation}
and introduce the power spectrum $P(k)$, defined as
\begin{equation}
\label{eq: power spectrum def}
    \langle \tilde{\delta}(\bfk)\,\tilde{\delta}(\bfk')\rangle=(2\pi)^3\,
    \delta_\mathrm{D}^{(3)}(\bfk+\bfk')\,P(k)\;,
\end{equation}
where $\delta_\mathrm{D}^{(3)}(\bfx)$ denotes the three-dimensional Dirac delta function.
The Wiener-Khinchin theorem states that the 2PCF and the power spectrum are Fourier-transform pairs, such that
\begin{equation}
    \xir(r)=\frac{1}{2\pi^2}\,\int P(k)\,j_0(kr) \, k^2 \,\mathrm{d}k\;,
\end{equation}
where $j_n(x)$ denotes the spherical Bessel function of order $n$.
An analogous relation holds true between the configuration- and Fourier-space three-point statistics
\begin{align}
\label{eq:3pcf from B}
    \zetar(r_{12}, r_{23}, r_{31}) &=
     \int  \,B(k_1, k_2, -|\bfk_1 + \bfk_2|) \, e^{i( \bfk_1 \cdot \bfr_{13} + \bfk_2 \cdot \bfr_{23} )} \frac{\mathrm{d}^3 k_1}{(2 \pi)^3} \frac{\mathrm{d}^3 k_2}{(2 \pi)^3} \,,
    \end{align}
where the bispectrum $B(k_1, k_2, k_3)$ is defined as \begin{equation}
\label{eq: bispectrum def}
    \langle \tilde{\delta}(\bfk_1)\, \tilde{\delta}(\bfk_2)\, \tilde{\delta}(\bfk_3) \rangle = (2\pi)^3\,\delta^{(3)}_\mathrm{D} (\bfk_1 + \bfk_2 + \bfk_3)\,B(k_1, k_2, k_3)  \; .
\end{equation}

\subsection{Redshift-space distortions}

In cosmology, we determine the distances of the observed objects from their redshifts assuming an unperturbed Friedmann-Lema\^itre-Robertson-Walker model universe.
This neglects several relativistic effects that alter the measured redshifts, the primary factor being the peculiar velocities of the objects. These velocities cause a displacement along the line of sight (LOS) between the reconstructed and actual comoving positions of the objects, $\bfs$ and $\bfx$. 
Throughout this work, we adopt the ``plane-parallel'' approximation, assuming that the LOS to any object is the same.  Using the $\zhat$ coordinate direction as the LOS, we can write the mapping from real to redshift space as follows:
\begin{equation}
\label{eq: rsd}
    \bfs= \bfx + \frac{v_{\parallel}}{a\,H} \, \zhat \equiv \bfx - w_{\parallel} \, \zhat\,,
\end{equation}
where $v_{\parallel}$ is the LOS component of the object's peculiar velocity (with respect to the observer), $a\,H$ is the conformal Hubble parameter, and, for convenience, we have defined a scaled velocity field, $w_{\parallel}$, in comoving distance units.

These so-called redshift-space distortions break statistical isotropy and imprint a LOS-dependence in the clustering properties of the objects.  Consequently, in redshift space, all previously introduced correlation functions depend on the orientation of the separation vectors (or wavevectors) relative to the LOS.  For example, the redshift-space 2PCF is not only a function of the magnitude of the pair separation vector, but also of the angle between the latter and the LOS.  Similarly, the 3PCF depends on the magnitude of the three separation vectors between the pairs in the triangle configuration, as well as on two angles that describe the orientation of the triangle with respect to the LOS: inspired by \cite{ScoCouFri9906} and \cite{Yankelevich2018}, we choose these as the polar angle $\theta$ between the LOS direction and the separation vector $\bfs_{12}$, and the azimuthal angle $\phi$ between the direction perpendicular to the plane of the triangle and the projection of the LOS on to the plane perpendicular to $\bfs_{12}$.

One can decompose the LOS-dependence of the two- and three-point correlation functions into a series of Legendre multipoles or spherical harmonics. For simplicity, in this paper we focus only on the monopole of the 3PCF, defined as the average over the two angles that describe the LOS orientation 
\begin{equation}
\label{eq: 3pcf monopole def}
    \zetas^0 (s_{12}, s_{23}, s_{31})= \frac{1}{4\pi} \int_{-1}^{1} \int_{0}^{2\pi}  \zetas(s_{12}, s_{23}, s_{31}, \theta, \phi) \, \mathrm{d}(\cos\theta) \, \mathrm{d}\phi \,.
\end{equation}
In order to distinguish quantities in redshift space from their real-space counterparts we use the subscript ``s''.

\subsection{Standard Perturbation Theory}
\label{sec: spt}

\subsubsection{Redshift-space bispectrum}
\label{sec:Bisp_SPT}

The statistical properties of the matter distribution can be predicted from an analytic approach called standard perturbation theory (SPT). In this context, the matter density field is assumed to be a perfect, pressureless fluid whose dynamics is fully captured by the Euler-Poisson system (see \cite{bernardeau2002} for a comprehensive review). While these assumptions are eventually violated by orbit crossing during the collapse into increasingly dense structures, SPT provides a reliable description on sufficiently large scales \cite[e.g.][]{Alkhanishvili2022}.
In this section, we report the expressions relevant for the computation of the 3PCF.

We begin by writing the density contrast in redshift space, using that the mapping from real to redshift space in Eq.~(\ref{eq: rsd}) conserves mass, i.e., $(1+\delta_\mathrm{s})\mathrm{d}^3 s = (1+\delta)\mathrm{d}^3 x $. In the plane-parallel approximation the Jacobian of the mapping is given by $J = 1 - \nabla_z\,w_{\parallel}$, and so we obtain the following representation in Fourier space:
\begin{equation}
  \label{eq: delta s from delta }
  \tilde{\delta}_\mathrm{s}(\bfk) = \int  e^{-i  \bfk\cdot \boldsymbol{x}} e^{i k_z\,w_{\parallel}} \Big[ \delta(\bfx)  + \nabla_z\,w_{\parallel}(\bfx)\Big] \, \mathrm{d^3}x\,.
\end{equation}
At this point, SPT proceeds by expanding the exponential factor resulting from the mapping for small values of $k_z\,w_{\parallel}$, which leads to the series
\begin{equation}
\label{eq: delta pt exp expansion}
    \tilde{\delta}_\mathrm{s}(\bfk)=\sum_{n=1}^\infty \int \delta_\mathrm{D}^{(3)}(\bfk - \bfk_{1\dots n}) \left[\tilde{\delta}(\bfk_1) + \nu^2 \tilde{\theta}(\bfk_1)\right] \frac{(\nu k)^{n-1}}{(n-1)!}\frac{\nu_2}{k_2}\tilde{\theta}(\bfk_2) \dots  \frac{\nu_n}{k_n}\tilde{\theta}(\bfk_n)\,\mathrm{d}^3 k_1  \dots \mathrm{d}^3 k_n\,,
\end{equation}
where $\nu_i \equiv k_{i,z}/k_i$ is the cosine of the angle between the wavevector $\bfk_i$ and the LOS, and $\theta(\bfx) \equiv \nabla \cdot \bfw(\bfx)$ is the velocity divergence.  Plugging this expression into Eq.~(\ref{eq: bispectrum def}) and using Wick's theorem under the assumption of Gaussian initial conditions we can express the redshift-space bispectrum in terms of auto- and cross-correlations between density and velocity fields. Up to fourth order in the perturbations, we get
\begin{eqnarray}
    \label{eq: SPT bispectrum}
    B_{\rms}^{\rm SPT}(\bfk_1, \bfk_2, \bfk_3) &=& B_{\delta \delta\delta}(k_1, k_2, k_3) +\bigg\{\bigg[  \nu^2_1 B_{\theta\delta\delta}(k_1, k_2, k_3)  + \nu ^2_1 \nu ^2_2 B_{\theta\theta\delta}(k_1, k_2, k_3) \nonumber\\
    && - \, \nu_3 k_3 \bigg\{ \frac{\nu _1}{k_1} \bigg[ P_{\delta\theta}(k_1) P_{\delta\delta}(k_2)  + \nu _1^2 P_{\theta\theta}(k_1)P_{\delta\delta}(k_2) + 2  \nu _2^2 P_{\delta\theta}(k_1)P_{\delta\theta}(k_2)  \nonumber \\
&&  + \, \nu _1^2  \left(2  \nu_2^2+  \nu _1^2\right) P_{\theta\theta}(k_1) P_{\delta\theta}(k_2)  + \nu _1^2 \nu _2^4 P_{\theta\theta}(k_1) P_{\theta\theta}(k_2) \bigg] \bigg\}  \nonumber \\
&&  + \, (1\leftrightarrow 2) \bigg] + \mathrm{cyc.} \bigg\} +  \nu^2_1 \nu^2_2 \nu^2_3 \, B_{\theta\theta\theta}(k_1, k_2, k_3)\,,
\end{eqnarray}
where $(1\leftrightarrow 2)$ indicates the exchange of $k_1$ and $k_2$, and $\mathrm{cyc.}$ the cyclic permutation over the three wavevectors. The real-space (auto and cross) power spectra and bispectra are defined in analogy with the definitions in~\cref{eq: power spectrum def,eq: bispectrum def}, e.g. $\langle  \tilde{\delta}(\bfk_1) \tilde{\theta} (\bfk_2) \rangle = (2\pi)^3 \, \delta_{\rm{D}}^{(3)}(\bfk_1 + \bfk_2) \, P_{\delta\theta}(k_1)$.

By recursively solving the Euler-Poisson system, SPT computes corrections to the linear solutions, $\tilde{\delta}_1$ and $\tilde{\theta}_1 = \tilde{\delta}_1/f$, where $f$ denotes the logarithmic growth rate of density perturbations. This results in a perturbative expansion,
\begin{equation}
\label{eq:delta pt redshift space}
\tilde{\delta}_s(\bfk ) = \sum_{n=1}^{\infty}  \int \delta_\mathrm{D}^{(3)} (\bfk - \bfk_{1\dots n} ) \,  Z_n(\bfk_1, \dots , \bfk_n) \,  \tilde\delta_1(\bfk_1) \dots \tilde\delta_1(\bfk_n) \,\mathrm{d}^3 k_1 \,  \dots \mathrm{d}^3 k_n  \,,
\end{equation}
where the corrections of order $n$ scale as $(\tilde{\delta}_1)^n$, and the kernel functions $Z_n(\bfk_1,\ldots,\bfk_n)$ describe the coupling of different scales due to non-linearities in the equations of motion and the real- to redshift-space mapping. In an Einstein-de Sitter background (with matter density parameter $\Omega_\mathrm{m} = 1$ and vanishing cosmological constant), the time dependence of the perturbative solutions is fully specified by the linear growth factor $D_1$. More precisely, at $n^\mathrm{th}$ order, it is given by the factor $(D_1)^n$. In other, more generic backgrounds, this scaling holds true as long as the ratio $\Omega_\mathrm{m}/f^2$ is close to unity. This is satisfied to excellent accuracy in $\Lambda$CDM \cite{bernardeau2002}, allowing us to absorb all the time dependence in the definition of $\tilde{\delta}_1$ (to simplify notation we do not write it explicitly).  Using Eq.~(\ref{eq:delta pt redshift space}), while still keeping only terms up to fourth order in $\tilde{\delta}_1$, we can simplify Eq.~(\ref{eq: SPT bispectrum}), which yields the usual SPT expression for the tree-level bispectrum in redshift space:
\begin{equation}
  \label{eq: spt bispectrum kernels}
  B_{\rm s}^{\rm SPT}(\bfk_1, \bfk_2, \bfk_3) = 2 \, Z_1(\bfk_1) \, Z_1(\bfk_2) \, Z_2(\bfk_1, \bfk_2) \, \pl(k_1) \, \pl(k_2) + \mathrm{cyc.}\,,
\end{equation}
Here, $\pl$ is the linear matter power spectrum and the two kernel functions required for the evaluation of this expression are given in Eqs.~(\ref{eq:Z1}) and (\ref{eq:Z2}).

\subsubsection{Infrared resummation}
\label{sec:IRresum}

It has long been recognised \cite[see, e.g.][]{MeiWhiPea9904,EisSeoWhi0708} that large-scale flows lead to a broadening and damping of the Baryon Acoustic Oscillation (BAO) in the 2PCF, an effect that is only poorly captured by SPT \cite{CroSco0801}.  This is because the displacements generated by long-wavelength perturbations affect the amplitude of the BAO feature by an amount comparable to the feature itself, implying that their perturbative expansion up to finite order leads to severe inaccuracies.  It was shown in \cite{SenZal1502} that these long-wavelength displacements can be treated non-perturbatively by exactly resumming their contributions to all orders, a procedure known as infrared (IR) resummation, which we will use to enhance the conventional SPT predictions.

Since the long-wavelength displacements leave the broadband shape of the power spectrum, $P_{\rm nw}$, invariant, the net effect on the linear power spectrum can be represented by damping its wiggly component\footnote{There is no unique method for isolating the oscillating part, $P_{\rm w}$, from the smooth part, $P_{\rm nw}$.  In this work we follow the method based on Gaussian filtering outlined in Appendix A of \cite{VlaSelYat1603}.} \cite{Baldauf2015, BlaGarIva1607}, $P_{\rm w}$:
\begin{equation}
\label{eq: IR ps}
    P_{\rm lin, IR}(k) = P_{\mathrm{nw}}(k) + e^{-k^2 \, \Sigma^2} \, P_\mathrm{w}(k)\,.
\end{equation}
The dispersion scale, $\Sigma$, relevant for the strength of the damping, is given by
\begin{equation}
\label{eq: IR sigma}
   \Sigma^2 = \frac{1}{6\pi^2}\int_0^{k_s} P_{\mathrm{nw}}(q)\left[1-j_0\left(q\,l_{\rm BAO}\right)+2j_2\left(q\,l_{\rm BAO}\right)\right] \mathrm{d}q\,,
\end{equation}
where $l_{\rm BAO}= 110\, h^{-1}\mathrm{Mpc}$ corresponds to the BAO scale.  The integral cutoff, $k_s$, serves to separate the long- and short-wavelength regimes and should in principle increase with $k$, i.e., $k_s = \epsilon\,k$ with fixed $\epsilon \ll 1$.  In practice, it is typically set to the fixed value $k_s = 0.2\,h\,\mathrm{Mpc}^{-1}$, which we do here as well.  As shown in \cite{IvaSib1807}, the leading order (LO) IR resummed bispectrum is similarly obtained by exchanging the linear power spectrum in the SPT expression (Eq.~\ref{eq: spt bispectrum kernels}) for the IR resummed one in Eq.~(\ref{eq: IR sigma}).  In redshift space, the additional displacements along the LOS direction produce an anisotropic broadening of the BAO feature, which leads to a LOS dependence in $\Sigma$ \cite{IvaSib1807}.  For the sake of simplicity, we ignore this complication and compute the SPT bispectrum in redshift space nonetheless with Eq.~(\ref{eq: IR ps}), which only introduces a negligible error in the monopole (with respect to the LOS).

\subsubsection{From the SPT bispectrum to the 3PCF}
\label{sec:bisp_to_3pcf}

The monopole of the 3PCF, $\zetas^0(s_{12},s_{23},s_{31})$, can be computed from the monopole of the bispectrum\footnote{The monopole of the bispectrum is defined analogously to Eq.~(\ref{eq: 3pcf monopole def}).}, $B_{\rm s}^0(k_1,k_2,k_3)$, via inverse Fourier transformation.  However, in practice, we express $\zetas^0$ in terms of two sidelengths and the angle between them, and expand the angular dependence in a series of Legendre polynomials,
  \begin{equation}
    \label{eq:zetas_expansion}
    \zetas^0(s_{12},s_{23},\hat{\bfs}_{12} \cdot \hat{\bfs}_{23}) = \sum_{L=0}^{\infty} \zeta_{\mathrm{s},L}^0(s_{12},s_{23}) \, {\cal L}_L(\hat{\bfs}_{12} \cdot \hat{\bfs}_{23})\,.
  \end{equation}
It can be shown that the coefficients $\zeta_{\mathrm{s},L}^0$ are related to the coefficients of an analogous expansion of the bispectrum monopole as follows \cite{SleEis1503}:
  \begin{equation}
    \label{eq: multipoles FT}
    \zeta^0_{\mathrm{s},L} (s_{12}, s_{23}) = (-1)^{L} 
    \int 
    \frac{k_1^2 \, k_2^2}{4\pi^4} \, B^0_{\mathrm{s},L}(k_1, k_2) \,
    j_L(k_1 s_{12})\,  j_L(k_2 s_{23}) \, \mathrm{d} k_1 \, \mathrm{d} k_2 \,.
  \end{equation}
  To model $\zeta_{\mathrm{s},L}^0$ using SPT, we first evaluate the coefficients from the SPT bispectrum monopole (obtained from Eq.~(\ref{eq: spt bispectrum kernels}) in combination with the IR resummation procedure of Section~\ref{sec:IRresum}),
  \begin{equation}
    \label{eq: bispectrum decomposition}
    B^0_{\mathrm{s},L}(k_1,k_2)=\frac{2 L+1}{2}  \int_{-1}^{1}  B^0_{\rm s} (k_{1}, k_2, \hat{\bfk}_1 \cdot \hat{\bfk}_2) \, \mathcal{L}_L(\hat{\bfk}_1 \cdot \hat{\bfk}_2) \, \mathrm{d}(\hat{\bfk}_1 \cdot \hat{\bfk}_2) \,,
  \end{equation}
  on a dense, logarithmically spaced grid of $k_1$ and $k_2$ values. Following \citep{umeh2021, Guidi2023}, we then apply the 2D-FFTLog algorithm by \cite{fang2021} (an extension of the original FFTLog approach introduced by \cite{hamilton2000}) to compute the remaining two-dimensional Bessel integral in Eq.~(\ref{eq: multipoles FT}).  The algorithm requires specification of various inputs, such as the integration ranges, the number of sampling points, and zero padding.  Our methodology and choices are thoroughly explained and validated in Appendix~\ref{app: 2dfftlog}.

\subsection{The streaming model} 
\label{sec: sm}

\subsubsection{General expressions}
\label{sec:SM_general}

Without expanding the mapping from real to redshift space as done in SPT, a more general, non-perturbative, relationship can be derived between correlation functions in real and redshift space. This relationship is known as the \textit{streaming model} and was first introduced and studied for the 2PCF by \cite{peebles1980, fisher1994}. As shown by \cite{scoccimarro2004}, the streaming model expression for the 2PCF follows only from mass conservation and the coordinate transformation in Eq.~(\ref{eq: rsd}), giving
\begin{equation}
\label{eq: 2pcf sm}
    1+ \xis(\bfs_{ij}) = \int  [1+\xir(r_{ij})] \, \mathcal{P}^{(2)}(w_{ij \parallel} | \bfr_{ij}) \, \mathrm{d}w_{ij\parallel}\,,
\end{equation}
where $\bfr_{ij}= \bfs_{ij} + w_{ij\parallel}\,\zhat$ with $w_{ij\parallel}$ the LOS component of the peculiar velocity difference between two points separated by $\bfr_{ij}$. The central quantity is the pairwise velocity probability density function (PDF), $\mathcal{P}^{(2)}(w_{ij \parallel} | \bfr_{ij})$, which maps pairs at separation $r_{ij\parallel}$ to separation $s_{ij\parallel}$ according to their relative velocity $-a\,H\,w_{ij\parallel}$.

Simulations show that the PDF exhibits a strong scale dependence: while for $r_{ij} \lesssim 1\,h^{-1}\mathrm{Mpc}$ and $r_{ij} \gtrsim 50\,h^{-1}\,\mathrm{Mpc}$ the PDF is predominantly symmetric, on intermediate scales it is skewed towards $w_{ij\parallel} < 0$ due to an increased probability of finding pairs that coherently fall into the same structures  (see, e.g., \cite{scoccimarro2004, kuruvilla2018, cuesta2020}). Its peak is typically close to $w_{ij\parallel} = 0$, indicating that the majority of pairs have uncorrelated velocities and thus do not belong to the same over- or underdense regions, where coherent (large-scale) flows would induce correlations. Furthermore, on all scales, the PDF has pronounced tails, such that it always deviates significantly from a Gaussian distribution.

The streaming model was generalised for the first time to higher-order statistics, in particular $n$-point correlation functions, by \cite{kuruvilla2020}. They showed that the 3PCF satisfies a relation analogous to the 2PCF,
\begin{equation}
\label{eq:3pcf sm}
\begin{split}
    1 + \xi_\rms(\bfs_{12}) + \xi_\rms(\bfs_{23}) + \xi_\rms(\bfs_{31}) + \zetas(\triangles) 
     =  \int  \big[1+\xi(r_{12}) +\xi(r_{23}) +\xi(r_{31})+\zeta(\triangler) \big] \\
     \times \, \mathcal{P}^{(3)}(w_{12 \parallel}, w_{23 \parallel} | \triangler) \, \mathrm{d}w_{12\parallel} \, \mathrm{d}w_{23\parallel} \,,
\end{split}
\end{equation}
but now involving the distribution of the LOS velocity differences in triplet configurations, $\mathcal{P}^{(3)}(w_{12 \parallel}, w_{23 \parallel} | \triangler)$. The notation $\triangler$ ($\triangles$) indicates the real- (redshift-)space configuration of the triplet.
Analogously to pairs, most triplets are not found within the same structures, such that the triplewise velocity PDF is uni-modal with a peak close to ($w_{12\parallel}, w_{23\parallel}) = (0,0)\,h^{-1}$Mpc, which was verified using simulation data \cite{kuruvilla2020}. It further shares all the scale-dependent properties noted above for ${\cal P}^{(2)}$, meaning that it displays skewness towards $w_{12\parallel} < 0$ or $w_{23\parallel} < 0$ depending on the size of $r_{12\parallel}$ and $r_{23\parallel}$, as well as extended tails. Since one object in the triplet necessarily appears twice in any two pairs (here, by choice, object `2'), the two velocity differences $w_{12\parallel}$ and $w_{23\parallel}$ are correlated, such that the PDF never appears symmetric in the $w_{12\parallel}-w_{23\parallel}$ plane. At large separations, their correlation coefficient approaches $-1/2$ \cite{kuruvilla2020}, as expected when drawing random LOS velocities at each of the three positions of the triplet.

\subsubsection{The Gaussian and Laplace streaming model}
\label{sec: gsm}

In the regime where the non-linearity of the mapping from real to redshift space becomes relevant, SPT breaks down and one can expect more accurate predictions from the streaming model. Evaluation of the streaming model, however, requires knowledge about the functional form of the pair- and triplewise LOS velocity PDFs, which is difficult to obtain. Measuring the PDFs from simulations is possible in principle, but due to their dependence on either pair separation or triangle configuration as well as orientation with respect to the LOS, it is computationally rather challenging.

Even without detailed knowledge of the PDFs, we can still advance our understanding of when the non-linearity in the mapping becomes significant.  To do this, we make simple assumptions about the shapes of the PDFs, using scale-dependent Gaussian and bivariate Gaussian distributions as our starting point.  This is motivated by the fact that on scales where skewness is less important, the central part of the PDFs tends to be roughly Gaussian \cite{scoccimarro2004,reid2011,kuruvilla2020}.  Indeed, for the 2PCF monopole and quadrupole, references
\cite{reid2011,wang2013} showed that this so-called Gaussian streaming model is accurate to within a few percent for scales $\gtrsim 20\,h^{-1}\mathrm{Mpc}$.  However, \cite{cuesta2020} argues that the agreement on these scales is driven mostly by an accurate description of the first and second moments, regardless of the exact shape of the PDF.  To see if this also holds true for the 3PCF, we compare the Gaussian streaming model against predictions from Laplace PDFs (known in one dimension as the \textit{double exponential} distribution).  Like Gaussians, Laplace distributions are defined by their first two moments, but have more extended tails, making them a better fit for the distribution of observed or simulated LOS peculiar velocities \cite[e.g.][]{Pee7604,She9604}.  Any differences between the Gaussian and Laplace predictions would therefore highlight the importance of the detailed shape of the velocity PDF.

The Gaussian and Laplace PDFs can be defined for the pairwise ($d=2$) and triplewise ($d=3$) cases as follows:
  \begin{align}
    {\cal P}_{\rm G}^{(d)}(\bfW) &\equiv \frac{1}{\sqrt{(2\pi)^{d-1}\,\mathrm{det}\,\boldsymbol{\mathsf{C}}}} \, \exp{\left[-\frac{1}{2} \left( \bfW - \langle \bfW \rangle\right)^T \, \boldsymbol{\mathsf{C}}^{-1} \, \left( \bfW - \langle \bfW \rangle\right)\right]} \,,     \label{eq:PDF_G} \\
    {\cal P}_{\rm L}^{(d)}(\bfW) &\equiv \frac{2}{\sqrt{(2\pi)^{d-1}\,\mathrm{det}\,\boldsymbol{\mathsf{C}}}} \, \left[-\frac{1}{2} \left( \bfW - \langle \bfW \rangle\right)^T \, \boldsymbol{\mathsf{C}}^{-1} \, \left( \bfW - \langle \bfW \rangle\right)\right]^{\frac{3-d}{4}} \nonumber \\
    &\hspace{1.5em} \times \, K_{\frac{3-d}{2}} \left[\sqrt{2 \left( \bfW - \langle \bfW \rangle\right)^T \, \boldsymbol{\mathsf{C}}^{-1} \, \left( \bfW - \langle \bfW \rangle\right)}\right] \,,     \label{eq:PDF_L} 
  \end{align}
  where $K_n$ denotes the $n^\mathrm{th}$ order modified Bessel function of the second kind.  The generic vector $\bfW$ stands for the LOS velocity difference and is either one- or two-dimensional (for $d=2$ or $d=3$, respectively), with corresponding mean vector $\langle \bfW \rangle$ and covariance matrix $\boldsymbol{\mathsf{C}}$.  We will refer to the streaming model with Gaussian or Laplace PDFs in the following simply as GSM and LSM, respectively.

To begin with,\footnote{An alternative option will be presented in Section~\ref{sec: moments}.} we evaluate $\langle \bfW \rangle$ and $\boldsymbol{\mathsf{C}}$ by using
the leading order (LO) SPT expressions, which we briefly review in the following (see also \cite{fisher1994,kuruvilla2020}).  Starting with the pairwise case, the mean LOS velocity at LO in SPT (including IR resummation) is given by 
\begin{equation}
  \label{eq: mean pairwise spt los}
  \langle w_{12 \parallel} \rangle_\mathrm{p} \lo \frac{f}{\pi^2 } \zhat \cdot {\hat{\bfr}}_{12} \int  \, k \, j_1(k r_{12}) \, \plir(k) \,\mathrm{d}k \equiv  \chi_{12} \, \bar{w}(r_{12})  \,,
\end{equation}
where the notation $\langle \cdot \rangle_{\mathrm{p}}$ indicates pair weighting and $\chi_{12} = \hat{\boldsymbol{z}} \cdot \hat{\bfr}_{12}$ is the cosine of the angle formed by the LOS and the pair separation vector.
For the LOS velocity dispersion one finds:
\begin{equation}
\label{eq: dispersion pairwise spt los}
\langle w_{12 \parallel}^2 \rangle_\mathrm{p} \lo 2 \Big[ \sigma_{v,\mathrm{lin}}^2 - \chi_{12}^2 \, \psi_r(r_{12}) - (1-\chi_{12}^2)\, \psi_p (r_{12}) \Big] \,,
\end{equation}
where $\sigma_{v,\mathrm{lin}}^2$ is the one-dimensional linear velocity dispersion 
\begin{equation}
\label{eq: sigmav^2}
    \langle w_{i\parallel}^2 \rangle \lo \sigma_{v,\mathrm{lin}}^2 \equiv \frac{f^2}{6 \pi^2} \int  \plir(k) \mathrm{d} k \,,
\end{equation}
and the correlation functions $\psi_r$ and $\psi_p$ are given by
\begin{eqnarray}
\label{eq: psi funcs}
    \psi_r (r_{12}) &=& \frac{f^2}{2 \pi^2} \int \left[ j_0(k\,r_{12}) - 2 \, \frac{j_1(k \, r_{12})}{k \, r_{12}} \right] \, \plir(k) \, \mathrm{d}k \,,\\
    \psi_p (r_{12}) &=&  \frac{f^2}{2 \pi^2} \int\frac{j_1(k\, r_{12})}{k\, r_{12}} \plir(k) \, \mathrm{d}k \,.
\end{eqnarray}
Zero-lag correlators, such as $\langle w_{i\parallel}^2\rangle$, do not only receive contribution from large-scale bulk flows, but also from virialised velocities within collapsed structures, as noted, e.g., in \cite{scoccimarro2004,reid2011}, which alters the large-scale limit of the pairwise dispersion.  To account for these contributions, we replace $\sigma_{v,\rm lin}^2$ in Eq.~(\ref{eq: dispersion pairwise spt los}) with its non-linear analogue $\sigma_{v}^2 \equiv \sigma_{v,\rm lin}^2 + C$, where $C$ is a constant whose value we will obtain by comparing to our measurements in Section~\ref{sec: moments}.

In the triplet configuration (with triplet weighting indicated here by the notation $\langle \cdot \rangle_{\triangle}$) the mean LOS velocity between objects `1' and `2' is modulated by the presence of the third object, leading to two additional terms compared to the pairwise case above:
\begin{eqnarray}
\label{eq: mean triplewise spt}
    \langle w_{12\parallel}  \rangle_\triangle
    & \lo &  \bar{w}(r_{12}) \, \chi_{12} - \frac{1}{2} \, \left[\bar{w}(r_{23}) \, \chi_{23}  + \bar{w}(r_{31}) \, \chi_{31} \right] \,.
\end{eqnarray}
This modulation does not occur at LO for the triplewise dispersion, which remains identical to the pairwise prediction, i.e.,
\begin{eqnarray}
\label{eq: dispersion triplewise spt}
   \langle w_{12\parallel}^2  \rangle_\triangle
   &\lo&  2 \Big[ \sigma_{v}^2 - \chi_{12}^2 \, \psi_r(r_{12}) - (1-\chi_{12}^2)\, \psi_p (r_{12}) \Big] \,,
\end{eqnarray}
where we have again exchanged the linear velocity dispersion $\sigma_{v,\rm lin}^2$ with its non-linear analogue.  Furthermore, the correlation of the two velocity differences $w_{12\parallel}$ and $w_{23\parallel}$ generates a non-zero mixed dispersion term, as anticipated earlier, which at LO is given by
\begin{eqnarray}
\label{eq: mixed second moment triplewise spt}
    \langle w_{12\parallel} w_{23\parallel}  \rangle_\triangle 
   & \lo &  \chi_{12}^2 \, \psi_r(r_{12}) + \left(1-\chi_{12}^2\right) \, \psi_p(r_{12}) + \chi_{23}^2 \, \psi_r(r_{23}) + \left(1-\chi_{23}^2\right) \, \psi_p(r_{23})  \nonumber\\
   & & - \; \chi_{31}^2 \, \psi_r(r_{31}) + \left(1-\chi_{31}^2\right) \, \psi_p(r_{31})  - \sigma_{v}^2 \,.
\end{eqnarray}
From these various components we can then define the mean and covariance that enter Eqs.~(\ref{eq:PDF_G}) and (\ref{eq:PDF_L}).  In the pairwise case we simply have $\langle \bfW \rangle = \langle w_{12\parallel} \rangle_{\rm p}$ and $\boldsymbol{\mathsf{C}} = \langle w_{12\parallel}^2 \rangle_{\rm p} - \langle w_{12\parallel} \rangle_{\rm p}^2$, while the triplewise moments are $\langle \bfW \rangle = \begin{pmatrix} \langle w_{12\parallel} \rangle_\triangle , & \langle w_{23\parallel}  \rangle_\triangle \end{pmatrix}$ and
\begin{equation}
\label{eq: triplewise covariance}
   \boldsymbol{\mathsf{C}}= \begin{pmatrix}
       \langle w_{12\parallel}^2  \rangle_\triangle - \langle w_{12\parallel}   \rangle_\triangle^2 & \langle w_{12\parallel} \, w_{23\parallel}   \rangle_\triangle - \langle w_{12\parallel}    \rangle_\triangle \langle w_{23\parallel}   \rangle_\triangle \\
       \langle w_{12\parallel} \, w_{23\parallel}   \rangle_\triangle - \langle w_{12\parallel}\rangle_\triangle \langle w_{23\parallel}  \rangle_\triangle & \langle w_{23\parallel}^2   \rangle_\triangle - \langle w_{23\parallel}  \rangle_\triangle^2
   \end{pmatrix} \,.
\end{equation}
Note that when evaluating the covariances at LO, the terms $\langle w_{12\parallel} \rangle_{\rm p}^2$, $\langle w_{12\parallel}\rangle_\triangle^2$ etc. vanish.

In order to compute the the Legendre coefficients of the 3PCF monopole, we rewrite Eq.~(\ref{eq:3pcf sm}) as a joint five-dimensional integral over the two velocity differences, the orientation of the LOS, as well as the angle between $\bfs_{12}$ and $\bfs_{23}$.  We solve this integral numerically using the \texttt{Cuhre} algorithm of the multidimensional integration library \texttt{CUBA} \cite{Hahn2005}.  We integrate the velocity differences over the interval $(-40, 40) \,h^{-1}\mathrm{Mpc}$ for the GSM and $(-100, 100)\,h^{-1}\mathrm{Mpc}$ for the LSM, since the Laplace distribution presents more prominent tails.

\begin{figure}[ht]
  \centering
  \includegraphics[width=\textwidth]{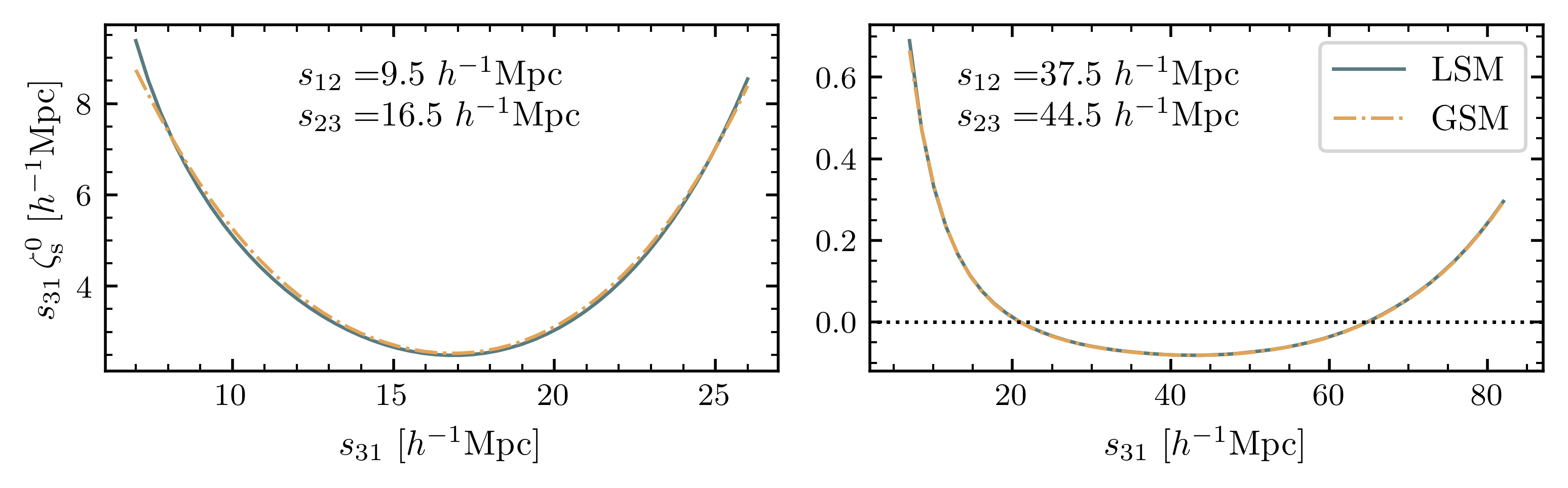}
  \caption{Monopole of the 3PCF in the Gaussian or Laplace streaming model obtained using velocity moments computed according to SPT at LO. Each panel represents a different configuration of fixed sides $s_{12}$ and $s_{23}$.}
  \label{Fig: LSM vs GSM}
\end{figure}

In Figure~\ref{Fig: LSM vs GSM}, we present a comparison between the three-point GSM and LSM using LO SPT predictions for the velocity moments.  The two panels illustrate different configurations, revealing that the differences in the 3PCF monopole between the two PDF prescriptions are negligible on large scales.  Only on scales smaller than $\sim 10\,h^{-1}\,\mathrm{Mpc}$, we obtain differences that are larger than 5\%.
As discussed above, this suggests that across a wide range of scales, the exact functional form of the LOS velocity PDF is not crucial for making accurate predictions of the 3PCF monopole.  This conclusion could still depend on whether skewness is taken into account or not, but measured skewness values from N-body simulations are negligible on large scales and only peak below $10\,h^{-1}\,\mathrm{Mpc}$ \cite{scoccimarro2004}.  
This item will be revisited in Section~\ref{sec: lsl}, where we will study the large-scale limit of the streaming model and show that, in this regime, the 3PCF is fully determined by the first two velocity moments, consistent with our findings here.

\section{Data and measurements}
\label{sec: data and measurements}
\subsection{N-body simulations}
\label{sec: data}

In order to assess the performance of the SPT and GSM models for the 3PCF, we compare their predictions against N-body simulations.  In this work we make use of the Quijote simulation suite \cite{villaescusa2019}; specifically, we pick 30 realisations at redshift $z=0$ from the $\Lambda$CDM set whose initial conditions were generated with second order Lagrangian perturbation theory.  Each of these simulations evolves a collection of $512^3$ dark-matter particles in a periodic cubic box of $1\,h^{-1}\mathrm{Gpc}$ per side using the TreePM code \textsc{Gadget-III} \cite{springel2005}.  The fiducial cosmological parameters are given by the matter density parameter $\Omega_\mathrm{m} = 0.3175$, baryon density $\Omega_\mathrm{b} = 0.049$, dimensionless Hubble constant $h= 0.6711$, spectral index $n_\mathrm{s} = 0.9624$, and amplitude of density fluctuations $\sigma_8=0.834$, which are in good agreement with the Planck cosmic microwave background constraints \cite{Planck2018}.

\subsection{Three-point correlation function estimator}
\label{sec: encore}

We generate the dark-matter distribution in redshift space by adopting the plane-parallel approximation and displacing the particles along the $\hat{\boldsymbol{z}}$-direction of the simulation boxes according to Eq.~(\ref{eq: rsd}). 

Next, we concentrate on the 3PCF averaged over all orientations of the LOS, i.e., the monopole moment with respect to the LOS, $\zeta^0_{\rm s}(s_{12}, s_{23}, s_{31})$.  
As already demonstrated in Section~\ref{sec:bisp_to_3pcf}, we can express this function in terms of two triangle sides and the cosine of the angle between them and subsequently expand the angular dependence in a Legendre series.
This results in the following estimator for the $L^\mathrm{th}$ multipole coefficient:
\begin{equation}
  \label{eq:data.3pcfestimator}
  \hat{\zeta}^0_{\mathrm{s},L}(s_{12},s_{23}) = (-1)^L \frac{\sqrt{2L + 1}}{4\pi\,V} \int \!\!\mathrm{d}^3x \int_{s_{12}} \!\!\!\!\! \mathrm{d}^3s_1 \int_{s_{23}} \!\!\!\!\!\mathrm{d}^3s_2 \, \delta(\bfx) \, \delta(\bfx + \bfs_1) \, \delta(\bfx + \bfs_2) \, {\cal L}_L(\hat{\bfs}_1 \cdot \hat{\bfs}_2)\,,
\end{equation}
where ${\cal L}_L$ are the Legendre polynomials of order $L$, $V$ denotes the volume of the simulation snapshot, and the integrals over $\bfs_1$ and $\bfs_2$ are performed over spherical shells of bin-width $\Delta s$ centred on $s_{12}$ and $s_{23}$, respectively.  This decomposition of the 3PCF was first proposed in \cite{slepian2015b} who showed that the $\hat{\zeta}^0_{\mathrm{s},L}$ coefficients can be estimated without explicitly counting triplets of galaxies, meaning that the computation process is significantly more efficient  than for the original estimator of the full 3PCF monopole \cite{szapudi1998}.

For our measurements, we employ this multipole estimator as implemented in the public code \textsc{Encore}\footnote{\url{https://github.com/oliverphilcox/encore} } \cite{philcox2021encore}, which evaluates Eq.~(\ref{eq:data.3pcfestimator}) by comparing the dark-matter particle distribution to a corresponding random catalogue that models the (constant) mean density.  Measuring the multipole coefficients for the entire set of particles is still computationally expensive, which is why we randomly downsample the original distributions from $512^3$ to $128^3$ particles, while we use a random catalogue that is 10 times denser.  The random selection guarantees that the clustering properties are unchanged and in Appendix~\ref{app:subsampling} we demonstrate the convergence of our measurements by considering different subsampling factors, as well as different sizes of the random catalogue.  We then measure the first eleven multipole coefficients with bin centres ranging from $s_{\rm min} = 20\, h^{-1}\rm{Mpc}$ to $s_{\rm max} = 160\, h^{-1}\rm{Mpc}$, and a bin size of $7 \,h^{-1}\rm{Mpc}$.  Because of the symmetry of the estimator, we restrict ourselves to configurations that satisfy\footnote{We do not consider isosceles configurations, $s_{12} = s_{23}$ as \textsc{Encore} cannot handle them. As noted in \cite{philcox2021encore}, a more sophisticated estimator would have to be implemented to deal with overlapping bins.} $s_{12} < s_{23}$. We also measure the monopole of the 2PCF from the same subsampled catalogues using the \textsc{Encore} implementation of the Landy-Szalay estimator \cite{landy1993}. We use the same range of scales and binning as for our 3PCF measurements.

\begin{figure}[t]
    \centering
\includegraphics[width=1.\textwidth]{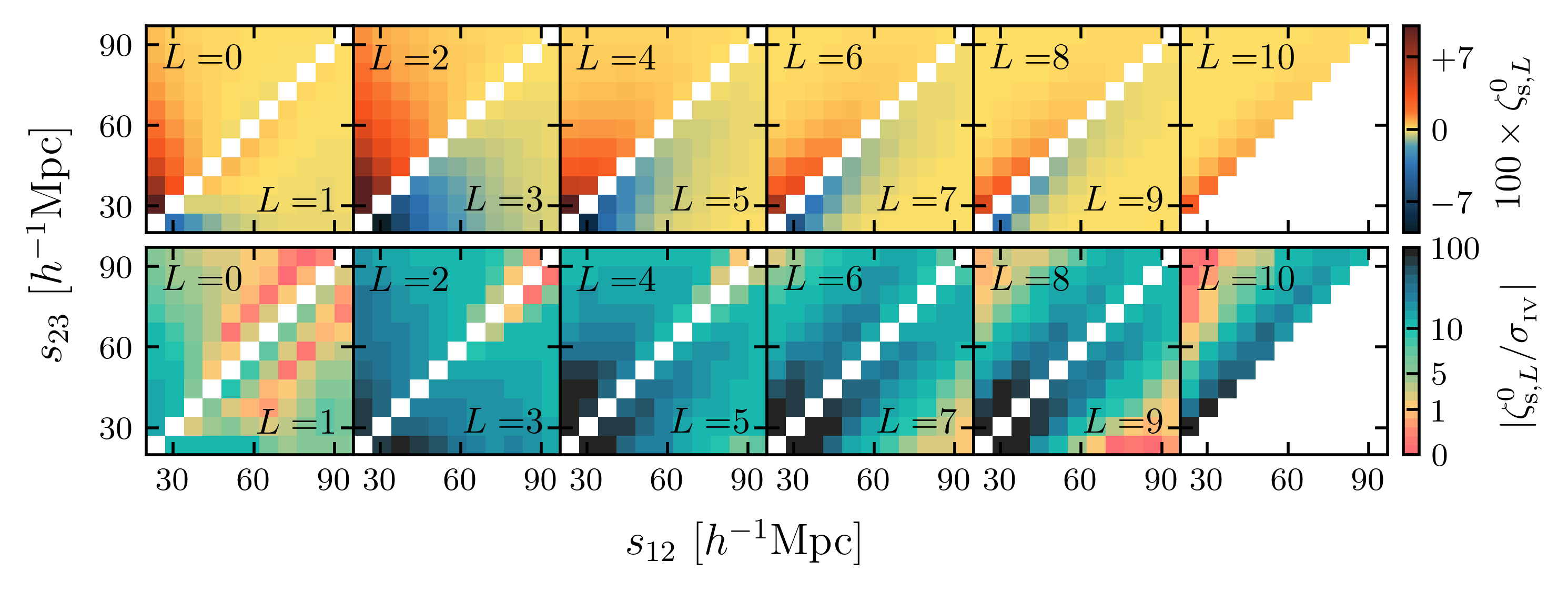}
    \caption{\textit{Top}: Isotropic 3PCF, averaged over 30 simulation measurements. \textit{Bottom}: Signal-to-noise ratio for each configuration bin assuming a Stage-IV survey. Each half-panel shows one Legendre multipole, as defined in Eq. (\ref{eq:zetas_expansion}). }
    \label{Fig: SN}
\end{figure}

In Figure~\ref{Fig: SN} (upper panels) we show the 3PCF measurements as a function of $s_{12}$ and $s_{23}$, averaged over the 30 realisations of the Quijote simulations.  Each panel displays two different multipoles above and below the diagonal, while the diagonal is left empty since we neglect isosceles configurations.  For all multipoles we notice an increasing signal towards smaller scales and towards configurations that are closer to being isosceles (closer to the diagonal in the plots).  The signal peaks in the multipole range from $L=2$ to $L=5$.  A similar picture also emerges for the signal-to-noise per bin, which is shown in the bottom panels of Figure~\ref{Fig: SN}. 
In this case, we divide the rms values of the measurements from
the Quijote simulations by the factor $\sqrt{8}$ to mimic
the statistical uncertainty corresponding to the typical volume
covered by a redshift bin in a Stage-IV survey, $V_{_\mathrm{IV}}\simeq 8 \,h^{-3} \,\mathrm{Gpc}^3$. We denote the resulting values with the symbol $\sigma_{_\mathrm{IV}}$.
While the amplitude of the signal-to-noise is suppressed for the first two multipoles, it is well above 10 for all following multipoles and grows towards smaller scales. We further notice that as the multipole number increases, the signal-to-noise ratios become increasingly concentrated in nearly isosceles configurations.  This is because for these configurations the averages over $\hat{\bfs}_{12} \cdot \hat{\bfs}_{23}$ involve triangle configurations where the third side ranges from $|s_{12} - s_{23}|$ to $s_{12} + s_{23}$, thus making them sensitive to the small-scale regime.  In this regime, non-linear couplings generate 3PCF contributions that depend on higher and higher powers of $\hat{\bfs}_{12} \cdot \hat{\bfs}_{23}$, which, in turn, furnish the signal that is measured by the higher-order multipoles. The resulting significant signal-to-noise ratios suggest (ignoring cross-correlations between the multipole measurements) that a full extraction of the 3PCF information may require measuring an extended range of multipoles.

To ease the comparison with theoretical predictions, we find it useful in some cases to combine the multipole measurements into a \textit{resummed} 3PCF,
\begin{equation}
\label{eq: zeta resummed}
{\zeta}^0_{\rms, (L_{\rm{max}})} (s_{12},s_{23},s_{31}) = \sum\limits_{L=0}^{L_{\rm{max}}} \frac{\sqrt{2 L +1} }{4\pi} (-1)^L \, \zeta^0_{\rms,L} (s_{12},s_{23}) \mathcal{L}_L\left(\frac{s_{31}^2-s_{12}^2-s_{23}^2}{2}\right)\,,
\end{equation}
 where $L_{\rm max}$ indicates the truncation order.  At finite truncation order, the resummed 3PCF is not necessarily close to the true 3PCF, especially given what we observed for the measurements presented above. In comparison with theory predictions, we therefore evaluate the theory multipole coefficients and then resum them as in Eq.~(\ref{eq: zeta resummed}).

\subsection{Velocity moments}
\label{sec: moments}

As we have seen in Section~\ref{sec: sm}, the central quantities in the GSM (LSM) are the mean velocities and dispersions, which fully characterise the Gaussian (Laplace) pair- or triplewise velocity PDFs.  In this section, we present measurements of these moments from the same 30 realisations of the Quijote simulations and compare them to the LO SPT predictions.

\subsubsection{Pair weighting}
\label{sec:pairweighting}

For the mean pairwise velocity, it is sufficient to consider the radial component $\langle w_{12 \rm r} \rangle_{\rm p} \equiv \langle \boldsymbol{w}_{12} \cdot \hat{\bfr} \rangle_{\rm p}$, since statistical isotropy implies that $\langle \bfw_{12} \rangle_{\rm p}$ must be aligned with the direction of the pair separation vector $\hat{\bfr}$ \cite[see, e.g.,][]{MoninYaglom1975}.  On the same grounds, one can show that the pairwise velocity dispersion tensor can be fully captured by just two components: one that is parallel to the separation vector, in addition to a transverse one.  We define these two components as $\langle w_{12 \rm r}^2\rangle_{\rm p} \equiv \langle \left(\bfw_{12} \cdot \hat{\bfr}\right)^2\rangle_{\rm p}$ and $\langle w_{12 \rm t}^2\rangle_p \equiv 1/2\,\langle \left[\bfw_{12} - \left(\bfw_{12} \cdot \hat{\bfr}\right)\hat{\bfr}\right]^2\rangle_{\rm p}$, respectively, and measure them along with the radial mean velocity in bins of $2\,h^{-1}\mathrm{Mpc}$ for scales up to $r_\mathrm{max}=280 \, h^{-1} \mathrm{Mpc}$.  To improve computational efficiency, we randomly downsample the particle distribution in each case to $64^3$ objects after having verified that this produces converged results, in the same way as what we did for the 3PCF in Appendix \ref{app:subsampling}. 

\begin{figure}[t]
    \centering
\includegraphics[width=1.\textwidth]{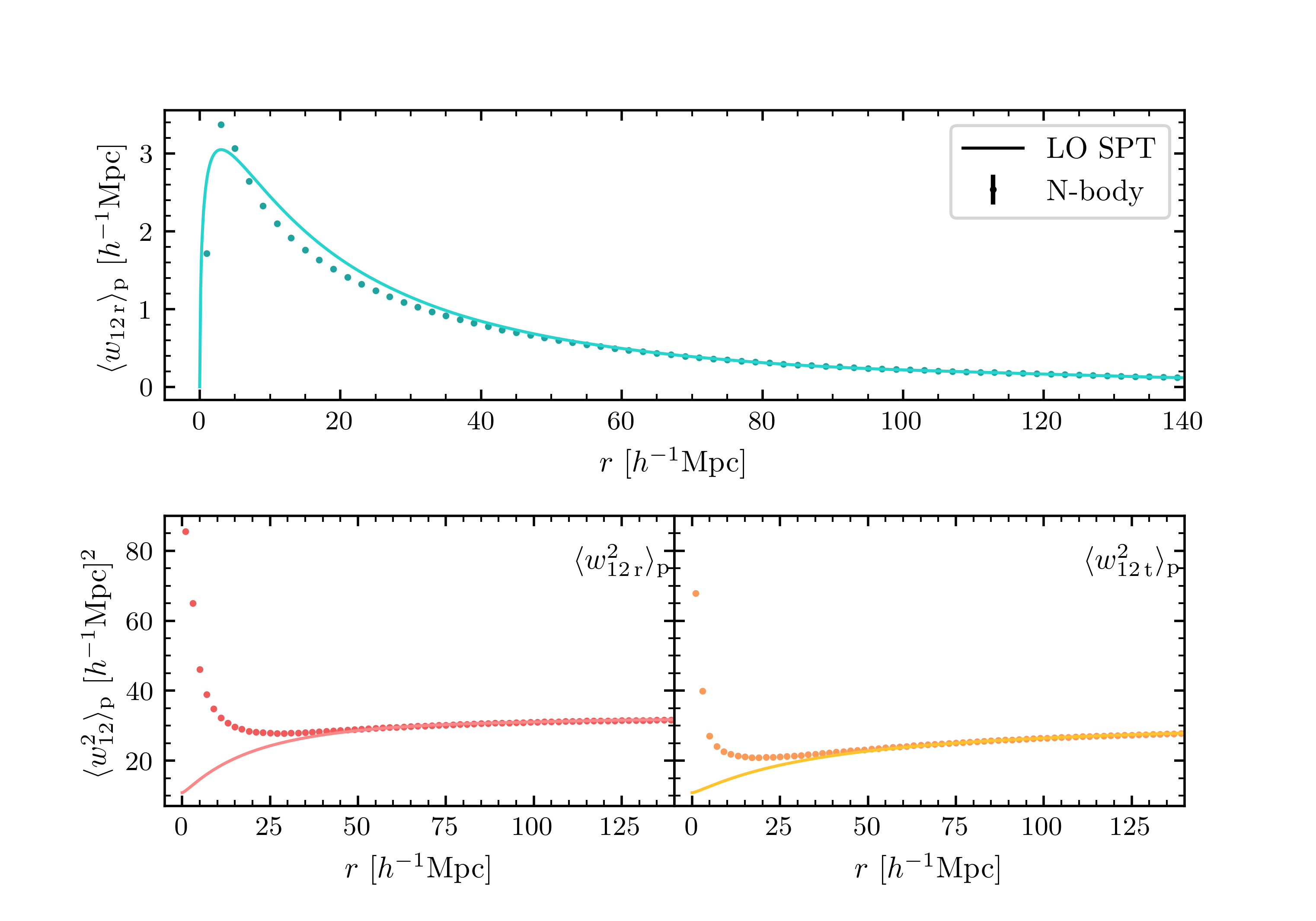}
    \caption{First two pairwise moments of the peculiar velocity differences, projected along radial and transverse directions and plotted as a function of separation scale. Measurements are depicted as symbols (with errorbars too small to be seen), whereas the predictions of LO SPT (including the non-linear shift of the dispersions) are shown by the lines.}
    \label{Fig: Moments pairwise}
\end{figure}

The measurements are shown in Figure~\ref{Fig: Moments pairwise} (dots) and are compared to the LO SPT predictions (lines), whose expressions are summarized in Appendix~\ref{app: velocity moments}.  We find that the agreement is generally better than 2\,\% on scales larger than $\sim 50\,h^{-1}\mathrm{Mpc}$, which is consistent with previous studies \cite[e.g.,][]{reid2011,cuesta2020}.  As expected, on smaller scales, non-linear evolution renders the LO SPT predictions increasingly discrepant, which is particularly evident in the velocity dispersions that become significantly underpredicted.  On large scales the dispersions approach a constant as they become dominated by the zero-lag correlator $\langle w_{i \rm r}(\bfx)^2 \rangle$ (where $i$ stands either for object `1' or `2').  As already discussed in Section~\ref{sec: gsm}, this quantity is highly sensitive to non-linearities, which leads to a constant offset with respect to its LO prediction in SPT, $\sigma_{v,\rm lin}^2$.  We have taken this into account by shifting $\sigma_{v,\rm lin}$ by the constant value $C$, and from a fit to the measurements we determine $C = 5.42 \, h^{-2}\mathrm{Mpc}^2$.

 \subsubsection{Triplet weighting}
\label{sec:tripletweighting}

\begin{figure}[t]
     \centering
     \begin{subfigure}[b]{0.45\textwidth}
         \centering
         \includegraphics[width=\textwidth]{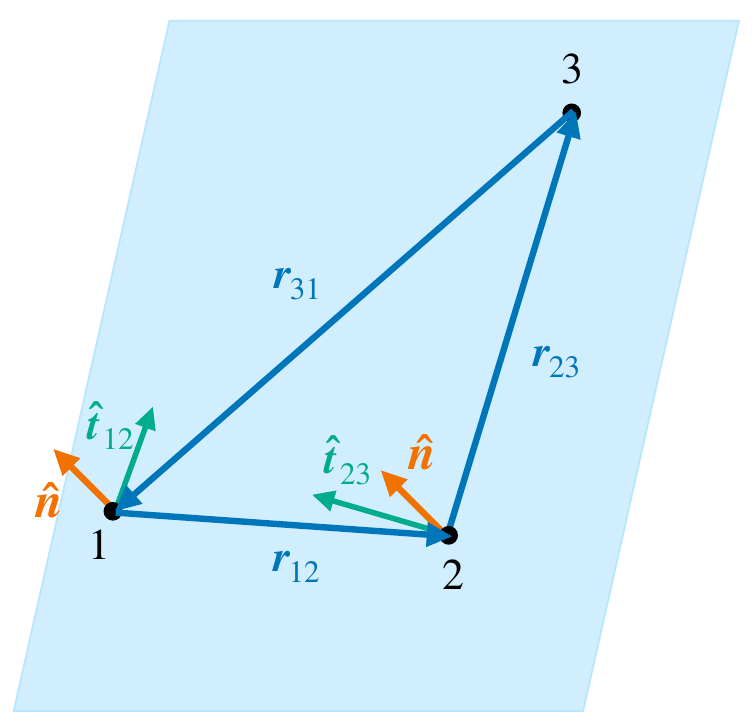}
         \caption{}
         \label{Fig: decomposition}
     \end{subfigure}
     \hfill
     \begin{subfigure}[b]{0.45\textwidth}
         \centering
         \includegraphics[width=\textwidth]{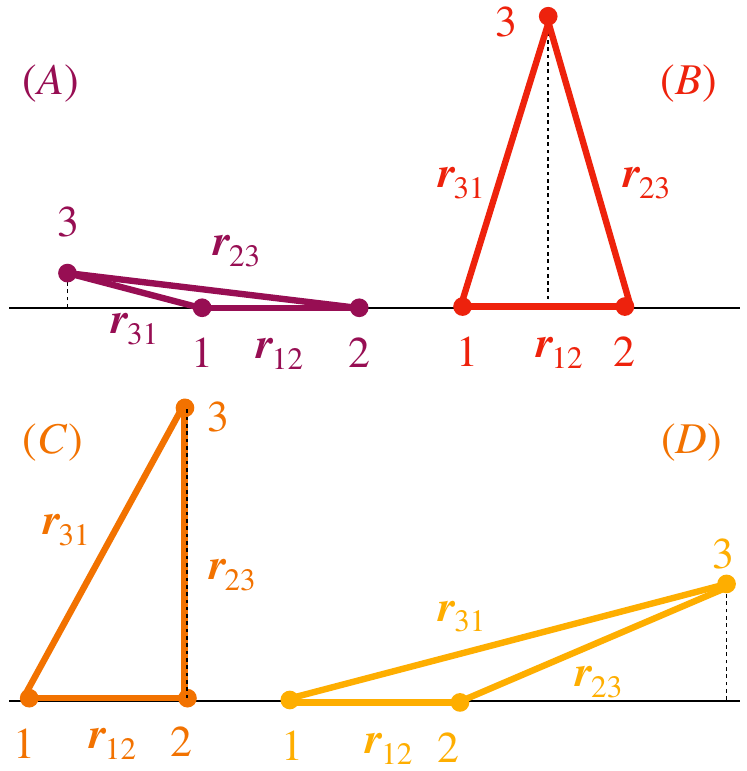}
         \caption{}
         \label{Fig: triangles}
     \end{subfigure}
  
     \caption{\textit{(a)}: Coordinate system adopted in the measurements of the triplewise velocity moments. For the triangle configuration defined by $1,2,3$, the triangle plane is indicated by the shaded region, and we define the vector normal to it as $\hat{\boldsymbol{n}}$ (orange), and the vectors $\hat{\boldsymbol{t}}_{12}$ and $\hat{\boldsymbol{t}}_{23}$ as $\hat{\boldsymbol{t}}_{12} = \hat{\boldsymbol{n}} \times \hat{\bfr}_{12}$ and
       $\hat{\boldsymbol{t}}_{23} = \hat{\boldsymbol{n}} \times \hat{\bfr}_{23}$ (in green).
       The moments of the velocity differences $\bfw_{12}$ and $\bfw_{23}$ are measured decomposed along $\{\hat{\bfr}_{12}, \hat{\boldsymbol{n}}, \hat{\boldsymbol{t}}_{12}\}$ and $\{\hat{\bfr}_{23}, \hat{\boldsymbol{n}}, \hat{\boldsymbol{t}}_{23}\}$, respectively.  \\
       \textit{(b)}: Different triangle configurations used in the discussion of the behaviour of the velocity moments.}
        \label{Fig:drawings}
\end{figure}

For triplet weighting, the mean velocity difference between any two of the three objects is no longer purely radial.  Instead, statistical isotropy implies that $\langle \bfw_{12} \rangle_\triangle$ has a contribution parallel to $\bfr_{12}$, but compared to the pairwise case it also acquires a second contribution parallel to $\bfr_{31}$.  We follow \cite{kuruvilla2020} and define an orthogonal coordinate system (see Figure \ref{Fig: decomposition} for a visual representation), consisting of the unit vector $\hat{\bfr}_{12}$, the vector perpendicular to the plane containing the triangle configuration, $\hat{\boldsymbol{n}} \equiv \hat{\bfr}_{12} \times \hat{\bfr}_{23}$, and a third vector within that plane, but transverse to $\hat{\bfr}_{12}$, giving $\hat{\boldsymbol{t}}_{12} \equiv \hat{\boldsymbol{n}} \times \hat{\bfr}_{12}$ (in the following we will refer to the latter two as \textit{normal} and \textit{transverse} directions, respectively).  We can then project the mean triplewise velocity differences into a component along $\hat{\bfr}_{12}$ and another along $\hat{\boldsymbol{t}}_{12}$, which we define as $\langle w_{12 \rm r}\rangle_\triangle \equiv \langle \bfw_{12} \cdot \hat{\bfr}_{12} \rangle_\triangle$ and $\langle w_{12 \rm t}\rangle_\triangle \equiv \langle \bfw_{12} \cdot \hat{\boldsymbol{t}}_{12} \rangle_\triangle$, respectively.  In the same manner we also take projections of $\langle \bfw_{23} \rangle_\triangle$, but using an analogous orthogonal coordinate system based on $\bfr_{23}$.

\begin{figure}[t]
    \centering
\includegraphics[width=1.\textwidth]{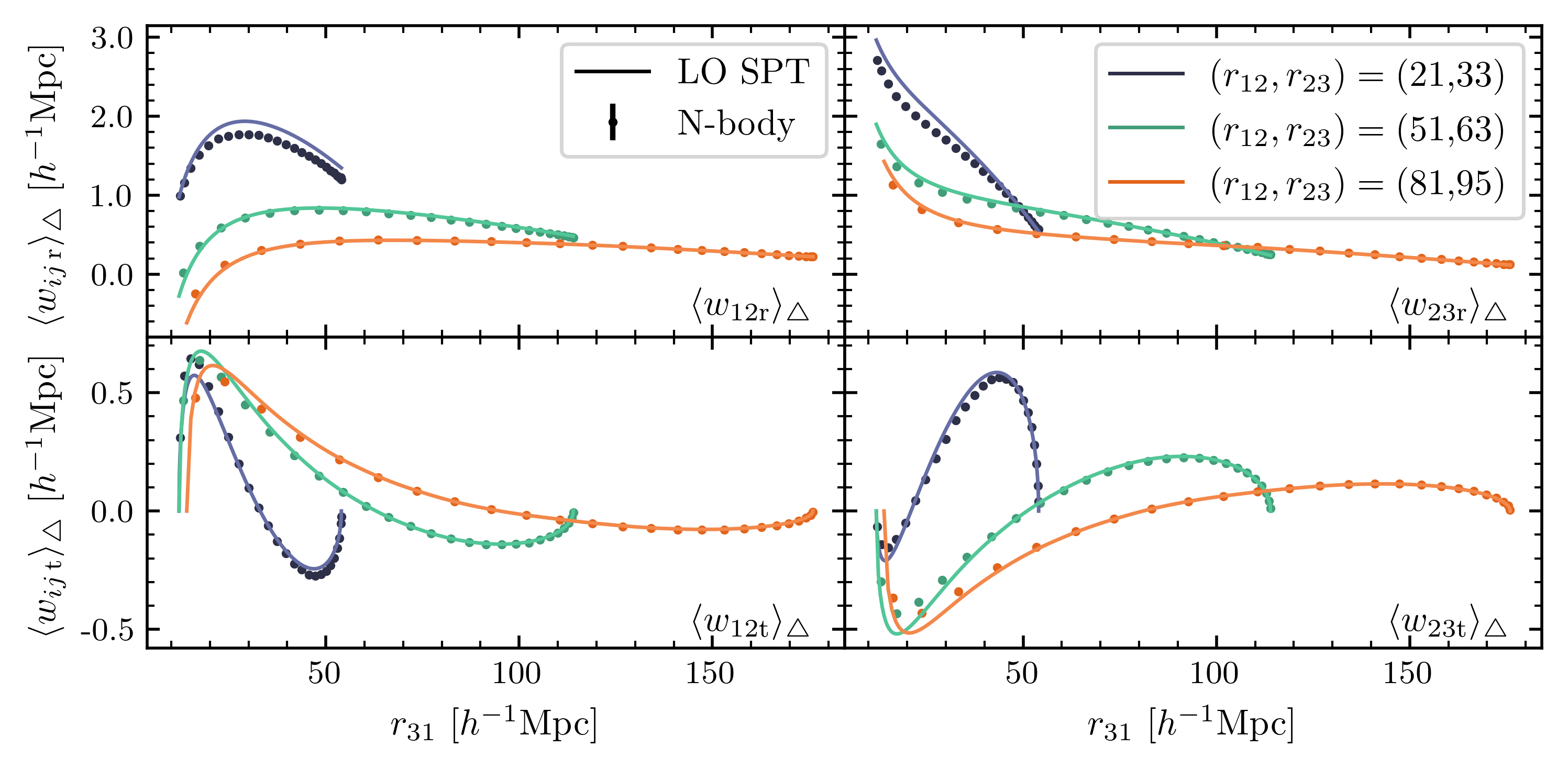}
\caption{Mean triplewise velocity differences between the galaxies of pair `12' (left column) and `23' (right column), projected along the radial and transverse directions. Different colours correspond to different configurations with fixed $r_{12}$ and $r_{23}$ (values
given in units of $h^{-1}\,\mathrm{Mpc}$), while symbols and lines represent the measurements and LO SPT predictions, respectively. Measurement errors are too small to be seen. To improve readability, we only plot every second data point. }
    \label{Fig: Moments trip first} 
\end{figure}

Each of these velocity components is a function of the triangle configuration $\{r_{12},r_{23},\chi = \arccos{\left(\bfr_{12} \cdot \bfr_{23}\right)}\}$ and we measure them on a regular grid up to scales of $r_\mathrm{max}=140 \, h^{-1} \mathrm{Mpc}$ in $r_{12}$ and $r_{23}$ with a bin size of $2 h^{-1}$Mpc, and 50 linearly spaced bins from 0 to $\pi$ in $\chi$.  The results are shown in Figure~\ref{Fig: Moments trip first} for three exemplary cases with fixed $r_{12}$ and $r_{23}$ and as a function of $r_{31} = \sqrt{r_{12}^2 + r_{23}^2 + 2 r_{12}\,r_{23}\,\cos{\chi}}$, which demonstrate how the mean velocity differences between two objects are modulated by the third object in the triplet.  Compared to the pairwise case, the radial component can either be diminished or enhanced, depending on the position of the third object relative to the other two.  In the examples shown, object `3' is outside\footnote{More precisely, by \textit{outside} we mean that the projection of object `$k$' onto the line passing through the pair `$ij$' does not lie between `$i$' and `$j$'.} the pair `12' for $r_{31} < \sqrt{r_{23}^2 - r_{12}^2}$ and $r_{31} > \sqrt{r_{23}^2 + r_{12}^2}$ (see triangle configurations A and D in Figure~\ref{Fig: triangles}, respectively), such that its different gravitational pull on the two objects reduces the velocity, $\langle w_{12 \rm r}\rangle_\triangle$, with which they approach each other.  This is most clearly seen for small $r_{31}$, i.e., at closest proximity to the pair (configuration A).  Conversely, if the third object is positioned between the other two,  it leads to an increase of the infall velocity.  This is the case  for object `1' with respect to the pair `23' if $r_{31} < \sqrt{r_{23}^2 + r_{12}^2}$ (as in triangle configurations A to C), leading to an evident boost in $\langle w_{23 \rm r}\rangle_\triangle$, but also for a range of $r_{31}$ values for $\langle w_{12 \rm r}\rangle_\triangle$ (for instance, configuration B).

The transverse component, $\langle w_{ij \rm t} \rangle_\triangle$, alternates between positive and negative values, and must vanish for symmetry reasons whenever the three objects are in a collinear configuration.  It also vanishes for isosceles configurations with $r_{ij}$ as the base (configuration B in Figure~\ref{Fig: triangles}), since the third particle exerts the same gravitational influence on the other two.  The sign  of the transverse mean velocity depends on which object of the pair `ij' is closer to the third: in the examples of Figure ~\ref{Fig: Moments trip first}, object `3' is initially closer to `1' (for small $r_{31}$), leading to a positive $\langle w_{12 \rm t} \rangle_\triangle$, whereas $\langle w_{23 \rm t} \rangle_\triangle$ must therefore have the opposite sign. Figure~\ref{Fig: Moments trip first} also shows the respective LO SPT predictions (see Appendix~\ref{app: velocity moments}), finding excellent agreement below $2\,\%$ provided that all three scales are $\gtrsim 50\,h^{-1}\mathrm{Mpc}$, which is consistent with the results presented for the mean pairwise velocity.

\begin{figure}[t]
    \centering
    \includegraphics[width=1.\textwidth]{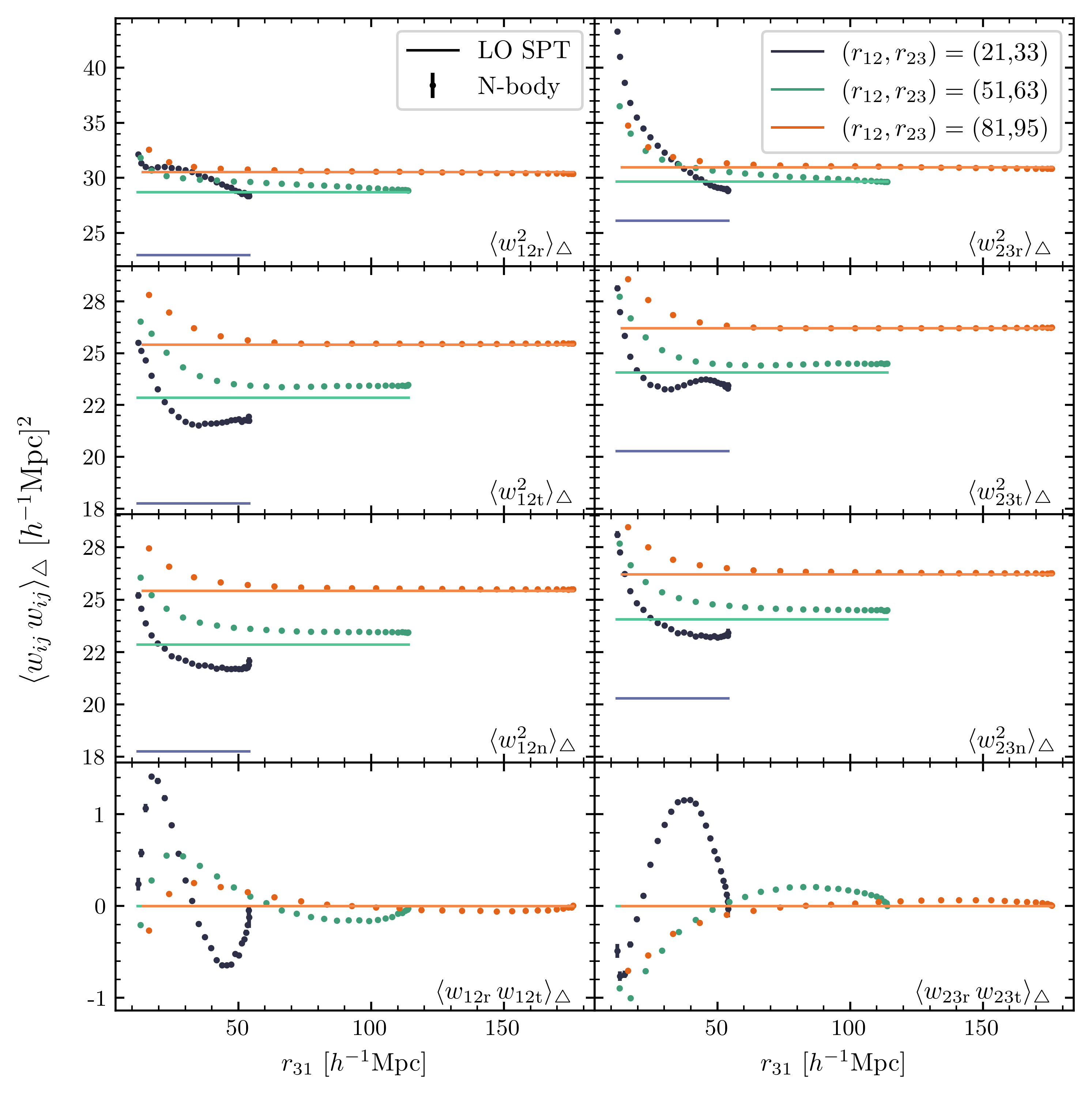}
    \caption{Same as Figure~\ref{Fig: Moments trip first}, but for the triplewise velocity dispersions. Each row depicts a different component of the dispersion tensor, Eq.~(\ref{eq:disp12_tensor_triplet}). 
    The SPT values for the average of the squared terms include the non-perturbative correction $C$ introduced in Section~\ref{sec: gsm}.
    Note that the LO SPT predictions for $\langle w_{ij \mathrm{r}} w_{ij \mathrm{t}} \rangle_\Delta$ vanish, so all solid lines overlap. }
    \label{Fig: Moments trip second}
\end{figure}

The dispersion tensor for velocity differences in triplets also has a more complex structure than its pairwise analogue.  From the invariance under rotations and reflections, as demanded by statistical isotropy, we can derive the following general form for the dispersion of the `12' pair (see, for instance, the argument presented in \cite{MoninYaglom1975}):
  \begin{align}
    \label{eq:disp12_tensor_triplet}
    \langle w_{12,i}\,w_{12,j} \rangle_\triangle = \; &\langle w_{12\rm{n}}^2  \rangle_\triangle\,\delta^{\rm K}_{ij} + \left[ \langle w_{12\rm{r}}^2  \rangle_\triangle -  \langle w_{12\rm{n}}^2  \rangle_\triangle\right] \hat{\bfr}_{12,i}\,\hat{\bfr}_{12,j} +  \left[ \langle w_{12\rm{t}}^2  \rangle_\triangle -  \langle w_{12\rm{n}}^2  \rangle_\triangle\right] \hat{\boldsymbol{t}}_{12,i}\,\hat{\boldsymbol{t}}_{12,j} \nonumber \\ &+ \frac{\langle w_{12\rm{r}}w_{12\rm{t}}  \rangle_\triangle}{2} \left(\hat{\bfr}_{12,i}\,\hat{\boldsymbol{t}}_{12,j} + \hat{\bfr}_{12,j}\,\hat{\boldsymbol{t}}_{12,i}\right)\,,
  \end{align}
  where $i,j$ here denote the entries of the velocity vectors, and $\delta^{\rm K}_{ij}$ indicates the Kronecker delta. Eq.~(\ref{eq:disp12_tensor_triplet}) implies that the dispersion tensor 
  receives four different contributions generated by different components of the velocity differences:
the purely radial or transverse components in the plane of the triangle configuration, $\langle w_{12\rm{r}}^2  \rangle_\triangle$ and $\langle w_{12\rm{t}}^2  \rangle_\triangle$, the mixed radial-transverse component $\langle w_{12\rm{r}}w_{12\rm{t}}  \rangle_\triangle$, and finally a component that is normal to the plane, $\langle w_{12\rm{n}}^2  \rangle_\triangle$, all of which are functions of the triangle configuration $\{r_{12},r_{23},r_{31}\}$.  Mixed components between the normal direction and any direction within the triangle plane must vanish because of symmetry, while for collinear triangle configurations, the form of the triplewise dispersion tensor must reduce to the pairwise case.  This means that the normal and transverse components have to become identical, $\langle w_{12\rm{n}}^2  \rangle_\triangle = \langle w_{12\rm{t}}^2  \rangle_\triangle$, and the radial-transverse part has to vanish.  An analogous form holds for the dispersion tensor of the `23' pair, with a separate set of four component functions, which we define with respect to the coordinate system $\{\hat{\bfr}_{23},\hat{\boldsymbol{t}}_{23},\hat{\boldsymbol{n}}\}$.  Finally, the same reasoning also applies to the mixed dispersion tensor and using a combination of the two sets of coordinate vectors, we obtain (suppressing the arguments of the component functions):
  \begin{align}
    \label{eq:disp12-23_tensor_triplet}
    \langle w_{12,i}\,w_{23,j} \rangle_\triangle = \; &\langle w_{12\rm{n}} \, w_{23\rm{n}}  \rangle_\triangle\,\delta^{\rm K}_{ij} + \left[\langle w_{12\rm{r}} \, w_{23\rm{r}}  \rangle_\triangle - \left(\hat{\bfr}_{12}\cdot \hat{\bfr}_{23}\right)\langle w_{12\rm{n}} \, w_{23\rm{n}}  \rangle_\triangle\right]\, \hat{\bfr}_{12,i}\,\hat{\bfr}_{23,j}\nonumber \\ 
    &+ \left[\langle w_{12\rm{t}} \, w_{23\rm{t}}  \rangle_\triangle - \left(\hat{\boldsymbol{t}}_{12}\cdot \hat{\boldsymbol{t}}_{23}\right)\langle w_{12\rm{n}} \, w_{23\rm{n}}  \rangle_\triangle\right]\, \hat{\boldsymbol{t}}_{12,i}\,\hat{\boldsymbol{t}}_{23,j} \nonumber \\
    &+ \langle w_{12\rm{r}} \, w_{23\rm{t}}  \rangle_\triangle\,\hat{\bfr}_{12,i}\,\hat{\boldsymbol{t}}_{23,j} + \langle w_{12\rm{t}} \, w_{23\rm{r}}  \rangle_\triangle\,\hat{\boldsymbol{t}}_{12,i}\,\hat{\bfr}_{23,j}\,,
  \end{align}
  which gives rise to one more radial-transverse component.  We note that both, Eq.~(\ref{eq:disp12_tensor_triplet}) and Eq.~(\ref{eq:disp12-23_tensor_triplet}), contain contributions
  that were not taken into account in \cite{kuruvilla2020} as the authors
  only considered the terms that do not vanish in SPT at LO.

\begin{figure}[t]
    \centering
    \includegraphics[width=1.\textwidth]{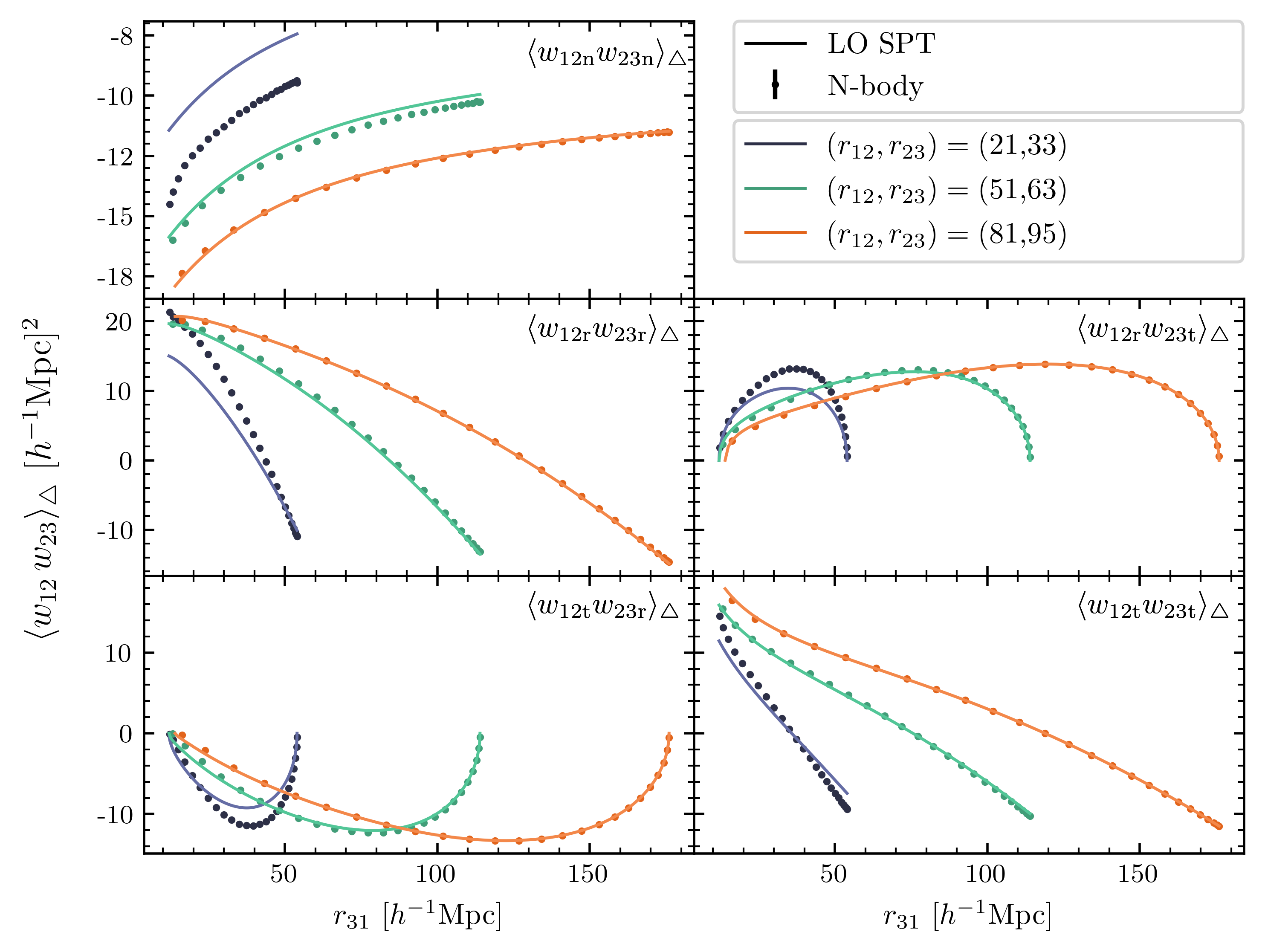}
    \caption{Same as Figure~\ref{Fig: Moments trip first}, but for the mixed triplewise dispersions between $w_{12}$ and $w_{23}$. Each panel depicts a different component of the dispersion tensor, Eq.~(\ref{eq:disp12-23_tensor_triplet}).}
    \label{Fig: Moments trip second mixed}
\end{figure}

In Figure~\ref{Fig: Moments trip second} we show the independent components of the $\langle w_{12,i}\,w_{12,j} \rangle_\triangle$ and $\langle w_{23,i}\,w_{23,j} \rangle_\triangle$ tensors for the same three configurations as before.  Due to the fixed pair-separation scale between the objects for which the velocity dispersion has been computed, the plot mainly visualises the impact of the third object in the triplet.  If this object is far away from the pair (large $r_{31}$) the measured values approach those of the pairwise case.  In close proximity, however, the third object acts to increase the velocity dispersion, irrespective of being positioned outside or within the other two objects, though we do note a bigger effect in the latter situation.  The mixed dispersion tensor has no analogue in the pairwise case, but from the measurement shown in Figure~\ref{Fig: Moments trip second mixed} we observe significant correlations between $\bfw_{12}$ and $\bfw_{23}$ in all five components.  The radial and transverse components are negatively correlated whenever the radial or transverse orientation align, i.e., $\hat{\bfr}_{12} \cdot \hat{\bfr}_{23} > 0$, $\hat{\boldsymbol{t}}_{12} \cdot \hat{\boldsymbol{t}}_{23} > 0$, and they switch sign if these orientations are anti-aligned.  For configurations in which these directions are perpendicular (as in triangle C in Figure~\ref{Fig: triangles}) the radial and transverse components vanish, while the mixed ones reach a maximum.

Like for pair weighting, the triplet-weighted dispersions are heavily affected by zero-lag contributions on all scales, which is why the measurements are not well modelled by the naive LO SPT predictions, which are given in Appendix~\ref{app: velocity moments}.  However, after shifting $\sigma_{v,\rm lin}$ by the same constant $C$ that we found by fitting the pairwise velocity dispersions, we obtain a per-cent level match for configurations with scales larger than $\sim 50 \, h^{-1}\mathrm{Mpc}$, as shown by the lines in Figures~\ref{Fig: Moments trip second} and \ref{Fig: Moments trip second mixed}.  Since the LO SPT model for the dispersions reduces to that of the pairwise case (which is why the predictions for the $\bfw_{12}$ and $\bfw_{23}$ are independent of $r_{31}$), deviations occur partly due to a mismatch of the pairwise dispersion, and partly because of the failure to capture the modulation by the third object.  The latter is especially obvious for small $r_{31}$, in which case the dispersions can be significantly underestimated.  Interestingly, the correlations between $\bfw_{12}$ and $\bfw_{23}$ are generally well modelled as a function of $r_{31}$, showing only slightly larger differences for smaller values of the pair separation scales $r_{12}$ and $r_{23}$ (blue data points).

\section{Comparing 3PCF models with N-body simulations}
\label{sec: results}

In this section, we compare the model predictions to the simulation measurements and analyse how well SPT and the GSM are able to capture the effects of RSDs on the 3PCF monopole.
  We consider two different implementations of the GSM: one that uses the LO SPT predictions for the velocity moments, and another that instead employs the measured velocity moments presented in the previous section.

\subsection{GSM for the 2PCF}
\label{sec:2pcfGSM}

\begin{figure}[t]
  \centering
  \includegraphics[width=1.\textwidth]{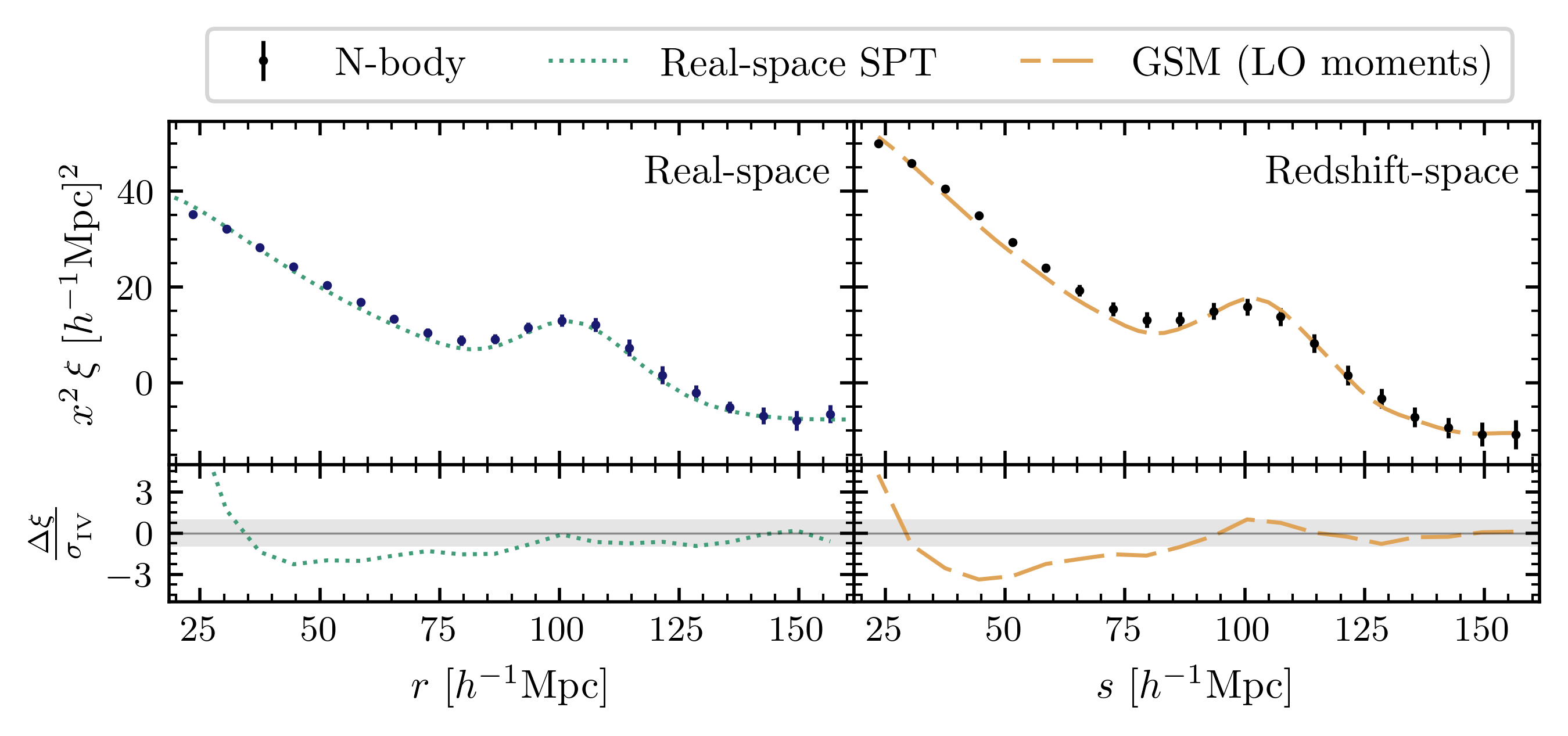}
  \caption{\textit{Top:} Mean real-space (left) and redshift-space (right) 2PCF measured from the Quijote simulations, compared to the IR-resummed linear-theory prediction (dotted) and the GSM based on the LO velocity moments (dashed). The 2PCF
  is scaled by either $r^2$ or $s^2$ in real- and redshift-space, respectively. Errorbars show the scatter $\sigma_{_\mathrm{IV}}$ expected in a Stage-IV survey.
  \textit{Bottom:} Normalised residuals. The shaded band marks deviations smaller than $\sigma_{_\mathrm{IV}}$.
}
  \label{Fig: model comparison jb gsm ptm 2pcf}
\end{figure} 

Since the 2PCF is a component of the 3PCF streaming model, let us begin by briefly reviewing the performance of the GSM for the 2PCF, before we discuss the 3PCF.  In Figure~\ref{Fig: model comparison jb gsm ptm 2pcf}, we show the mean 2PCF extracted from the simulations in real and redshift space, compared to the linear 2PCF (including IR resummation, see Section~\ref{sec:IRresum}) and the GSM, respectively.  The latter is computed using the linear 2PCF as input along with the LO predictions for the mean infall and velocity dispersions, shifted by the constant offset $C$.

For the adopted measurement uncertainties corresponding to a Stage-IV like volume, we find that the linear 2PCF accurately describes the real-space measurements within $1\sigma_{_\mathrm{IV}}$ (represented by the grey error band in the bottom panels) for scales $\gtrsim 90\,h^{-1}\,\mathrm{Mpc}$.  For smaller scales, the 2PCF is underpredicted by up to $3\sigma_{_\mathrm{IV}}$, and eventually overpredicted on even smaller scales.  The GSM with LO moments displays residuals that are very similar to the real-space case for scales larger than about $60\,h^{-1}\,\mathrm{Mpc}$.  On these larger scales, the primary source of the modelling discrepancy is therefore the real-space 2PCF.  The modelling of the velocity moments must accordingly be adequate, which is consistent with our comparison against the measured moments presented in Section~\ref{sec:pairweighting}.

\subsection{The (partially) resummed 3PCF}
\label{sec: res sm}
\begin{figure}[t]
  \centering
  \includegraphics[width=0.97\textwidth]{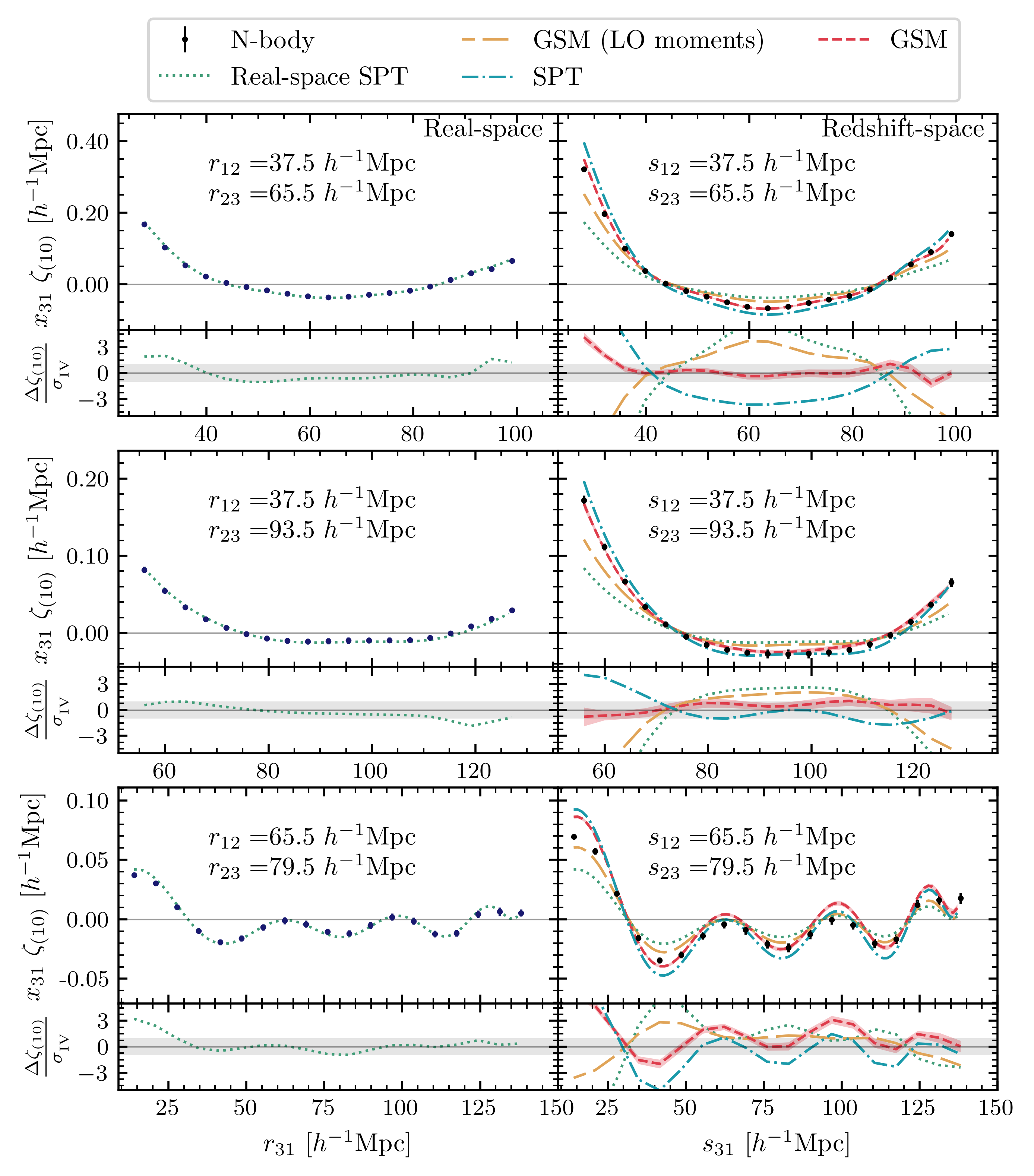}
  
  \caption{\textit{Left:} The resummed 3PCF (up to $L=10$) in real space from the Quijote simulations is compared to the tree-level SPT prediction for three different triangle configurations (the triangle sides indicate the bin centres). The bottom sub-panels show the corresponding normalised residuals. The grey shaded bands indicate deviations smaller than $1\sigma_{_\mathrm{IV}}$.  
    \textit{Right:} The resummed monopole of the redshift-space 3PCF extracted from the simulations is compared to different model predictions; 
    the dotted line shows the real-space model for comparison. The GSM model is surrounded by an error band resulting from a bootstrap resampling of the measured velocity moments.
    Note that the resummed 3PCF is multiplied by $r_{31}$ and $s_{31}$ in real- and redshift-space, respectively.
    }
  \label{Fig: model comparison jb gsm ptm}
\end{figure}

We now consider the ``resummed'' 3PCF  ${\zeta}^0_{(10)} $ obtained by limiting the sum over the multipoles in Eq.~(\ref{eq:zetas_expansion}) to $L\leq 10$.
As in the previous section, we first consider the real-space 3PCF, which is shown for three representative triangle configurations in the left panels of Figure~\ref{Fig: model comparison jb gsm ptm}.
The measurements are obtained using the same estimator as presented in Section~\ref{sec: encore} and for a total of 10 realisations (opposed to the 30 used in redshift space), from which the rms value has been computed and then rescaled to the volume corresponding to a typical redshift bin in a Stage-IV survey (see Section~\ref{sec: encore}).  The dotted lines, representing the tree-level SPT prediction (the explicit expressions are given in Appendix~\ref{sec:app.real-space-tree}), are an excellent fit to these measurements.  They show a level of agreement that is well within  $1\sigma_{_\mathrm{IV}}$ down to scales of $\sim 25\,h^{-1}\,\mathrm{Mpc}$, and we also find comparable results for other triangle configurations.  Any discrepancies between our models and the measurements in redshift space will therefore be dominated by the treatment of the real- to redshift-space mapping, and not the modelling of the real-space 3PCF.

The corresponding resummed measurements in redshift space are plotted in the right-hand panels of Figure~\ref{Fig: model comparison jb gsm ptm}, and the dashed yellow lines show the predictions of the GSM when evaluating all ingredients using LO SPT\footnote{The LO SPT predictions for the pair- and triplewise dispersions include the constant shift $C$, as discussed in Section~\ref{sec: moments}.}.  Compared to the real-space results, the level of agreement is significantly degraded, even on relatively large scales, $\gtrsim 60\,h^{-1}\,\mathrm{Mpc}$, where the GSM model appears to be almost identical to the real-space model (shown again, for reference, by the dotted lines) as already noted in \cite{kuruvilla2020}.  Given the good agreement in real space, as well as our previous observation that the isotropic 3PCF does not depend on the precise functional form of the triplewise velocity PDF (see Section~\ref{sec: gsm}), the discrepancy is likely due to insufficient modelling of the velocity moments.  This conclusion appears to be in disagreement with Section~\ref{sec: moments}, where we showed that the measured velocity moments are well described by the LO SPT predictions on scales larger than $40\,h^{-1}\,\mathrm{Mpc}$.  Moreover, as seen in Section~\ref{sec:2pcfGSM}, the LO SPT description of the moments is not a limiting factor for the 2PCF on these scales.  We will clarify this seemingly contradictory behaviour by exploring the large-scale limit of the streaming model in Section~\ref{sec: lsl}.

Let us now consider the performance of the GSM, presuming we had perfect knowledge of the velocity moments.  To do so, we evaluate the GSM by interpolating our measured pair- and triplewise velocity moments while still using the tree-level SPT predictions for the real-space 2PCF and 3PCF.  The results are shown as the dashed red lines in Figure~\ref{Fig: model comparison jb gsm ptm} and demonstrate a notable improvement compared to when the velocity moments are modelled through LO SPT.  The agreement with the measured 3PCF is now generally within $1\sigma_{_\mathrm{IV}}$ across a wide range of scales, although we still observe increasing discrepancies for the first configuration where $r_{31}$ drops below $\sim 40\,h^{-1}\,\mathrm{Mpc}$ (top panel), and in particular for the nearly isosceles configuration, which appears worse than for the GSM with moments from LO SPT (bottom panel).

In order to exclude the possibility that these discrepancies are driven by noise in the measured moments, we recomputed the GSM using 50 bootstrap samples of the moments that were obtained by averaging over 30 resamplings of the 30 measurements (allowing for repetitions).  
We then computed the scatter across the resulting 50 GSM predictions, which is shown as a shaded error band around the nominal model.  This error is at most a few percent and clearly does not account for the discrepancies, demonstrating that measurement noise is not significantly affecting our model predictions.  A bigger impact may result from the finite binsizes in $r_{12}$ and $r_{23}$, as well as the fixed sampling of the angle between the two separation vectors, which could lead to inaccuracies in the interpolation process.  This could in principle be easily overcome, but would require a much larger amount of computational resources, and so could not be easily tested here.

Finally, Figure~\ref{Fig: model comparison jb gsm ptm} also shows the 3PCF predictions using the Fourier-transformed tree-level SPT model for the bispectrum (dot-dashed, blue lines), as explained in Section~\ref{sec:bisp_to_3pcf}.  While being an improvement over the GSM with LO SPT moments for the configuration with one small and one large side (middle panel), it displays a similar level of agreement for the others.  We note, in particular, that the full SPT model fails to accurately recover the 3PCF when one, or especially two triangle sides approach scales smaller than $\sim 60\,h^{-1}\,\mathrm{Mpc}$, and for these cases only the GSM with measured moments yields better results.  For the third (nearly isosceles) configuration shown in the figure, SPT and the GSM with measured moments both fail to accurately match the measurements.

\subsection{Multipoles of the 3PCF}
\label{sec: res model comparison}

\begin{figure}[t]
\centering
\includegraphics[width=1.\textwidth]{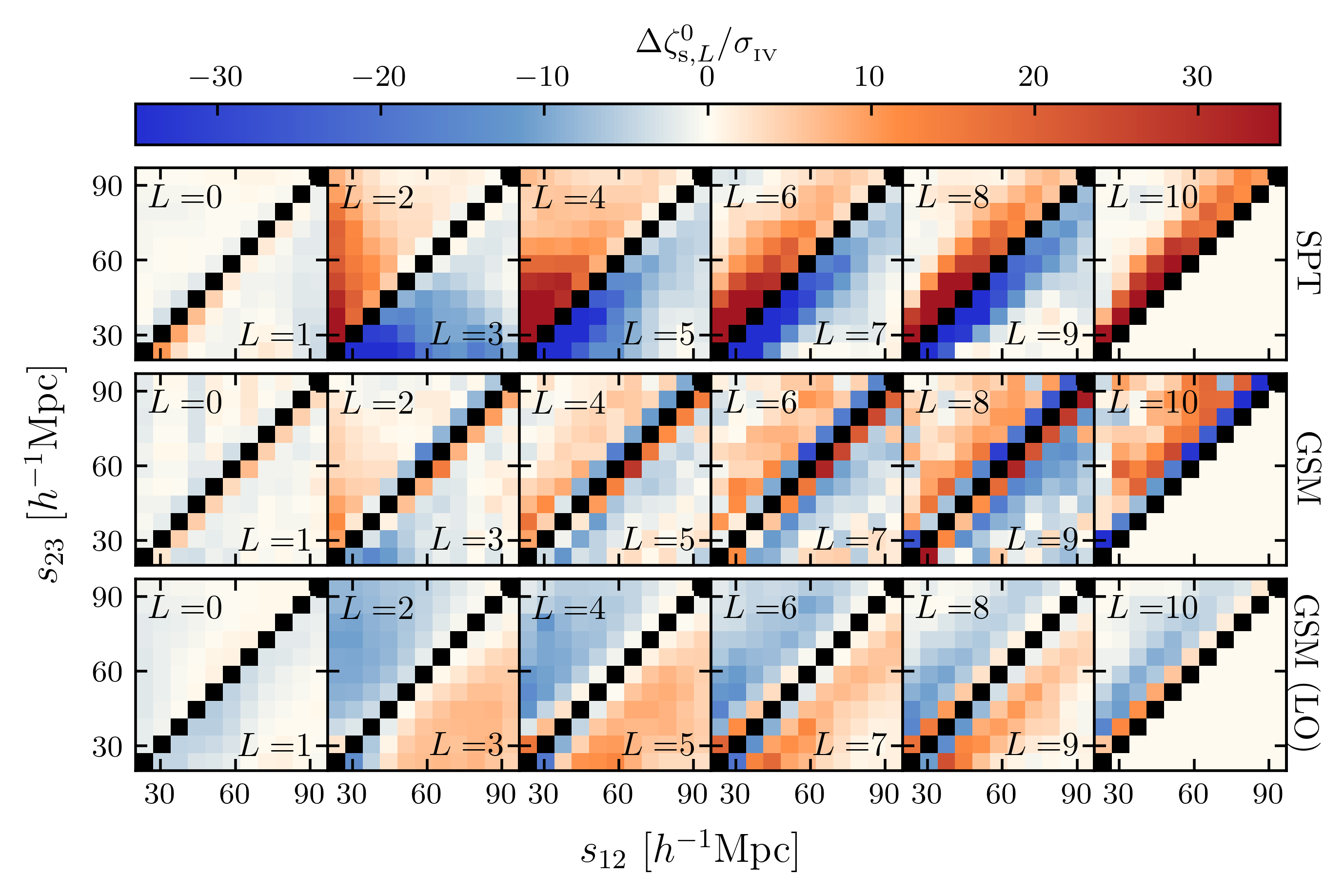}
    \caption{Differences between measurements and theoretical predictions of the isotropic 3PCF multipoles in units of standard deviations corresponding to a Stage-IV like volume.  Each panel displays a different multipole above and below the diagonal (exploiting the symmetry in $s_{12}$ and $s_{23}$), while each row shows a different  model. Namely, from top to bottom: SPT, GSM with moments measured from the simulations, and GSM with LO SPT predictions for the moments.}
    \label{Fig: model comparison spt gsm mm diff multipoles}
\end{figure}

In order to complement the comparison presented in the previous section, we now study how well the different models can describe the 3PCF multipoles directly.  The residuals between all individually measured multipoles and their corresponding model predictions (in units of standard deviations for the Stage-IV like volume) are plotted in Figure~\ref{Fig: model comparison spt gsm mm diff multipoles}.  All three models share a common feature: they recover the first two multipoles with roughly equal accuracy, within the $\sim 2\sigma_{_\mathrm{IV}}$ bounds.  However, more substantial differences arise in the higher-order multipoles, which are more relevant for extracting cosmological information since they contain the majority of the signal-to-noise (see Figure~\ref{Fig: SN}).  For these multipoles, we observe that the GSM with LO SPT moments (bottom row) consistently overpredicts the even multipoles and underpredicts the odd ones by a similar amount.  There is only a slight degradation when approaching smaller scales for $s_{12}$ and $s_{23}$.  On the other hand, the full SPT model (top row) displays the opposite behaviour, tending to overpredict the even multipoles and underpredict the odd ones.  Moreover, the failure of the model becomes more significant when the two separations are small, whereas its performance is somewhat better than the GSM with LO SPT moments for configurations where at least one side is larger than $\sim 60\,h^{-1}\,\mathrm{Mpc}$, which is consistent with our previous observations for the resummed 3PCF.  The GSM with measured moments (middle row) provides a significantly better description of the $L=2$ - $5$ multipoles compared to the two other models, which explains the improved behaviour in the resummed 3PCF.  However, it fails for nearly isosceles configurations as well as for the highest multipoles, where it yields results worse than for the GSM with LO SPT moments.  As noted above, this could be driven less by a limitation of the GSM itself, but more by interpolation errors of the measured velocity moments.  Since the higher multipoles quantify faster variations due to changes in the angle between $\bfs_{12}$ and $\bfs_{23}$, one can expect interpolation errors to have a stronger impact in these cases.

\section{The large-scale limit of the streaming model}
\label{sec: lsl}

In the previous sections, we observed that the GSM with LO SPT predictions for the velocity moments fails to accurately model the redshift-space 3PCF.  This happens even though the moments themselves are well-modelled on large scales, and the same approach works better for the 2PCF.  However, for small scales and nearly isosceles configurations, the GSM with LO SPT moments provides a better description than the full SPT model, which expands the mapping from real- to redshift-space perturbatively.  This is despite the streaming model not depending on the precise functional form of the PDF on the scales of interest for this work.  In this section, we will clarify these seemingly contradictory observations by evaluating the streaming model in the large-scale limit and explaining the conditions for a connection with SPT.

\subsection{Derivation}
\label{sec:LL_derivation}

What exactly do we mean by large-scale limit? Given that even moments of the pair- and triplewise PDFs do not vanish for large separations (see Section \ref{sec: moments}), we cannot simply apply a small-amplitude expansion in the density and velocity fluctuations.  However, as was noted in \cite{scoccimarro2004} for the 2PCF, for large separations that satisfy $s_{ij\parallel}^2 \gg \langle w_{ij\parallel}^2\rangle$, the integrations over the velocity PDFs in the streaming model are sharply peaked around $r_{ij\parallel} = s_{ij\parallel}$.
That implies that we can study the behaviour of the streaming model on these large scales by expanding real-space quantities around their redshift-space counterparts.  As we will show in the following, this will allow us to derive a local relationship between the real- and redshift-space quantities that can be matched (under certain additional assumptions) with SPT.

The derivation and comparison with SPT simplifies when expressing the three-point streaming model in a manifestly symmetric manner. We therefore consider all relevant quantities as functions of individual positions and velocities with respect to the observer, instead of pair separations and velocity differences, i.e.,
  \begin{eqnarray}
  \label{eq:list}
      & \xi(\bfr_{ij}) \rightarrow \xi(\bfr_i,\bfr_j)\,, & \quad
        {\cal P}^{(2)}(w_{ij\parallel}|\bfr_{ij}) \rightarrow {\cal P}^{(2)}(w_{i\parallel},w_{j\parallel}|\bfr_i,\bfr_j)\,, \\
      & \zeta(\bfr_{12},\bfr_{23}) \rightarrow \zeta(\bfr_1,\bfr_2,\bfr_3)\,, & \quad  {\cal P}^{(3)}(w_{12\parallel},w_{23\parallel}|\bfr_{12},\bfr_{23}) \rightarrow {\cal P}^{(3)}(w_{1\parallel},w_{2\parallel},w_{3\parallel}|\bfr_1,\bfr_2,\bfr_3)\,, \nonumber
  \end{eqnarray}
  and integrate over all three LOS velocities $w_{i\parallel}$. Statistical homogeneity guarantees that this formulation is fully equivalent with what was presented above and we can recover Eqs.~(\ref{eq: 2pcf sm}) and (\ref{eq:3pcf sm}) by integrating out one of the LOS velocities. In the large-separation limit, we can then expand the quantities listed in Eq.~(\ref{eq:list}) around each individual position $r_{i\parallel} = s_{i\parallel}$\footnote{Technically, the streaming model integrands only peak for pair separations $r_{ij\parallel} = s_{ij\parallel}$, which leaves freedom for a constant shift vector in the individual positions of the objects. However, by virtue of statistical homogeneity the final result cannot depend on this shift vector, which is why we ignore it to begin with.}, and starting from the 2PCF streaming model, we get
  \begin{align}
    \label{eq:2pcf_expansion}
    \xi_s(\bfs_i,\bfs_j) &\simeq \int \sum_{n=0}^N \sum_{\substack{m_1,m_2 \geq 0 \\ m_1+m_2 = n}} \frac{w_{i\parallel}^{m_1}\,w_{j\parallel}^{m_2}}{m_1!\,m_2!} \, \frac{\partial^{m_1}}{\partial s_{i\parallel}^{m_1}} \frac{\partial^{m_2}}{\partial s_{j\parallel}^{m_2}} \nonumber \\ & \hspace{2em} \times \, \Big\{\big[1 + \xi(\bfs_i,\bfs_j)\big] \, {\cal P}^{(2)}(w_{i\parallel},w_{j\parallel}|\bfs_i,\bfs_j)\Big\} \, \mathrm{d}w_{i\parallel} \mathrm{d}w_{j\parallel} - 1 \nonumber \\
    &= \xi(\bfs_i,\bfs_j) + \sum_{n=1}^N \sum_{\substack{m_1,m_2 \geq 0 \\ m_1+m_2 = n}} \frac{1}{m_1!\,m_2!} \, \frac{\partial^{m_1}}{\partial s_{i\parallel}^{m_1}} \frac{\partial^{m_2}}{\partial s_{j\parallel}^{m_2}}  \Big\{ \big[1 + \xi(\bfs_i,\bfs_j)\big] \, \mu^{(2)}_{m_1 m_2}(\bfs_i,\bfs_j)\Big\}\,,
  \end{align}
  where we have organised the multi-dimensional Taylor series such that it runs over an outer sum counting the total number of derivatives up to the expansion order $N$. In the second step we have also used the definition of the pairwise LOS velocity moments, i.e.
\begin{align}
\label{eq: pairwise moment def  pdf los}
  \mu_{m_1 m_2}^{(2)}(\bfs_i, \bfs_j)  &\equiv  \int  w_{i\parallel}^{m_1} w_{j\parallel}^{m_2} \, \mathcal{P}^{(2)}(w_{i\parallel},w_{j\parallel} | \bfr_i,\bfr_j)\,\mathrm{d}w_{i\parallel} \mathrm{d}w_{j\parallel} \\
 &\sst \frac{\langle w_{i\parallel}^{m_1} w_{j\parallel}^{m_2} \left(1 + \delta_i\right) \left(1 + \delta_j\right) \rangle}{\langle \left(1 + \delta_i\right) \left(1 + \delta_j\right) \rangle} \,, \label{eq:mu_pair}
\end{align}
with $\delta_i \equiv \delta(\bfs_i)$, and, in the second row, we have assumed the single-stream regime --i.e., that there is a well-defined single velocity at each location (with no velocity dispersion)-- indicated by the overscript SSt. 
Truncating the expansion in Eq.~(\ref{eq:2pcf_expansion}) at order $N=2$ and imposing statistical homogeneity, we recover what was originally derived for the 2PCF streaming model in \cite{scoccimarro2004}. 

Let us now turn to the 3PCF streaming model whose large-scale limit can be obtained in a completely analogous way. We expand the integrand in Eq.~(\ref{eq:3pcf sm}) and subtract the redshift-space 2PCFs using Eq.~(\ref{eq:2pcf_expansion}), which yields
\begin{align}
\label{eq: lsl general}
  \zetas(\bfs_1,\bfs_2,\bfs_3) &\simeq \zetar(\bfs_1,\bfs_2,\bfs_3) + \sum_{n=1}^N \sum_{\substack{m_1,m_2,m_3 \geq 0 \\ m_1+m_2+m_3=n}} \frac{1}{m_1!\,m_2!\,m_3!} \, \frac{\partial^{m_1}}{\partial s_{1\parallel}^{m_1}} \frac{\partial^{m_2}}{\partial s_{2\parallel}^{m_2}} \frac{\partial^{m_3}}{\partial s_{3\parallel}^{m_3}} \nonumber \\ &\times \Big\{\big[1 + \xirud + \xirdt + \xirtu + \zetar \big] \mu_{m_1 m_2 m_3}^{(3)}(\bfs_1,\bfs_2,\bfs_3)  - \left[1+ \xirud\right]\mu_{m_1 m_2}^{(2)}(\bfs_1, \bfs_2) \Big. \nonumber \\  & \Big. \hspace{2em} - \left[1+ \xirdt\right] \mu_{ m_2 m_3}^{(2)}(\bfs_2, \bfs_3) - \left[1+ \xirtu \right] \mu_{m_3 m_1}^{(2)}(\bfs_3, \bfs_1)   \Big\}\,,
\end{align}
where all the quantities in the square brackets are evaluated at the respective redshift-space positions, i.e., $\xirud =\xir (\bfs_1, \bfs_2)$ and $\zetar = \zetar(\bfs_1,\bfs_2,\bfs_3)$, and we have defined the triplewise LOS velocity moments as follows:
\begin{align}
\label{eq: triplewise moment def  pdf los}
\mu_{m_1 m_2 m_3}^{(3)}(\bfs_1,\bfs_2,\bfs_3)  &\equiv \int  w_{1\parallel}^{m_1} w_{2\parallel}^{m_2} w_{3\parallel}^{m_3} \, \mathcal{P}^{(3)}(w_{1\parallel}, w_{2\parallel}, w_{3\parallel} | \bfr_1,\bfr_2,\bfr_3)\,\mathrm{d}w_{1\parallel} \mathrm{d}w_{2\parallel} \mathrm{d}w_{3\parallel} \\
& \sst \frac{\langle w_{1\parallel}^{m_1} w_{2\parallel}^{m_2} w_{3\parallel}^{m_3} \left(1 + \delta_1\right) \left(1 + \delta_2\right) \left(1 + \delta_3\right)\rangle}{\langle \left(1 + \delta_1\right) \left(1 + \delta_2\right) \left(1 + \delta_3\right) \rangle}\,. \label{eq:mu_triple}
\end{align}
As for the 2PCF we note that the result in Eq.~(\ref{eq: lsl general}) is completely general in the regime where the large-scale limit is valid, no assumptions regarding the functional form of the velocity PDFs have been made.

\subsection{Connection to SPT}
\label{sec:LL_SPT}

The large-scale limit expansion clearly shows that the redshift-space corrections to the 3PCF are determined by the difference between the triplewise and pairwise velocity moments. This means that the 3PCF captures not only the correlation between LOS velocities at all three positions but also how pairwise infall, dispersion, etc. fluctuate more or less strongly depending on the overdensity at the third object's location in the triplet (as observed in Section~\ref{sec: moments}). These correlations arise from non-linearities and cannot be fully captured when evaluating the moments with LO SPT (up to quadratic order in perturbations). At this order, only the first two moments are finite (in the sense of not being infinitesimal), but the difference between the triple- and pairwise moments vanishes nonetheless, as can be verified using the expressions in Eqs.~(\ref{eq:mu_pair}) and (\ref{eq:mu_triple}). Therefore, under the assumption of LO SPT moments, we simply recover the real-space 3PCF in the large-scale limit. For the 2PCF, the situation is different: in the large-scale limit, it directly depends on the pairwise moments, so restriction to LO does not eliminate the effect from RSDs.

Expanding the expressions to fourth order in the perturbations, i.e. up to second order in the linear power spectrum, we can match the large-scale limit of the 3PCF streaming model term by term to the SPT result in Eq.~(\ref{eq: SPT bispectrum}). At this order, there are non-zero contributions from the first four moments, and, keeping only the finite terms, we find the following matching after Fourier transformation (FT):
\begin{itemize}
\item \textbf{First order moments ($n=1$):}
  \begin{equation}
    \label{eq: lsl n=1}
    \begin{split}
    \mathrm{FT} &\bigg[\frac{\partial}{\partial s_{1\parallel}}\bigg( \langle \delta_2 \delta_3 w_{1\parallel} \rangle + \langle \delta_1 \delta_2\rangle \langle\delta_3 w_{1\parallel}\rangle + \langle\delta_1 \delta_3 \rangle\langle \delta_2 w_{1\parallel}\rangle \bigg) + \mathrm{cyc.} \bigg] \\
                &\hspace{-1em}= \nu_1^2 B_{\theta\delta\delta} (\bfk_1, \bfk_2, -\bfk_1- \bfk_2) 
                   - \nu_3 k_3  \bigg[ \frac{\nu_1}{k_1} P_{\delta\theta}(k_1) P_{\delta\delta}(k_2) + (1 \leftrightarrow 2) \bigg] + \mathrm{cyc.}
    \end{split}
  \end{equation}
\item \textbf{Second order moments ($n=2$):}
  \begin{equation}
    \label{eq: lsl n=2}
    \begin{split}
      \mathrm{FT}  \bigg[ & \dfrac{\partial^2}{\partial s_{1\parallel}^2} \langle\delta_2 w_{\parallel 1} \rangle \langle\delta_{3} w_{\parallel 1}\rangle + \frac{\partial^2}{\partial s_{1\parallel }\partial s_{2\parallel}} ( \langle\delta_3 w_{\parallel 1} w_{\parallel 2} \rangle + \langle \delta_2 \delta_3 \rangle \langle w_{\parallel 1} w_{\parallel 2} \rangle  + \langle \delta_3 \delta_1 \rangle \langle w_{\parallel 1} w_{\parallel 2} \rangle \\
      & + \langle\delta_2  w_{\parallel 1} \rangle \langle\delta_{3}  w_{\parallel 2}\rangle +  \langle\delta_3 w_{\parallel 1} \rangle \langle\delta_{1} w_{\parallel 2}\rangle ) + \mathrm{cyc.}\bigg] = \nu ^2_1 \nu ^2_2 B_{\theta\theta\delta}(\bfk_1, \bfk_2, \bfk_3) \, \\
      &   -  \nu_3 k_3 \bigg\{ \frac{\nu _1}{k_1} \bigg[ \nu _1^2 P_{\theta\theta}(k_1)P_{\delta\delta}(k_2)  + 2  \nu _2^2 P_{\delta\theta}(k_1)P_{\delta\theta}(k_2) \bigg] + (1\leftrightarrow 2) \bigg\} + \mathrm{cyc.}
    \end{split}
  \end{equation}
\item \textbf{Third order moments ($n=3$):}
  \begin{equation}
    \label{eq: lsl n=3}
    \begin{split}
    \mathrm{FT} &\bigg[\dfrac{\partial^3}{\partial s_{1\parallel}^2 \partial s_{2\parallel}} \langle\delta_3 w_{\parallel 1}\rangle\langle w_{\parallel 1}w_{\parallel 2} \rangle + \dfrac{\partial^3}{\partial s_{1\parallel }\partial s_{2\parallel}^2} \langle\delta_3 w_{\parallel 2}\rangle\langle w_{\parallel 1}w_{\parallel 2} \rangle \\ 
    &+ \dfrac{\partial^3}{\partial s_{1\parallel }\partial s_{2\parallel}\partial s_{3\parallel}} \big( \langle w_{\parallel 1} w_{\parallel 2}w_{\parallel 3} \rangle 
    + \langle\delta_1 w_{\parallel 2}\rangle \langle w_{\parallel 1}w_{\parallel 3} \rangle + \langle\delta_1 w_{\parallel 3}\rangle \langle w_{\parallel 1}w_{\parallel 2} \rangle \big) + \mathrm{cyc.} \bigg] \\ 
    &\hspace{-1em}=\nu_1^2 \nu_2^2 \nu_3^2 B_{\theta\theta\theta}(\bfk_1, \bfk_2, \bfk_3) - \nu_3 k_3 \left[  \frac{\nu_1^3}{k_1}  \left(2 \nu_2^2+  \nu _1^2\right) P_{\theta\theta}(k_1) P_{\delta\theta}(k_2)  + (1 \leftrightarrow 2) \right]  + \mathrm{cyc.}
    \end{split}
  \end{equation}
\item \textbf{Fourth order moments ($n=4$):}
  \begin{equation}
    \label{eq: lsl n=4}
    \begin{split}
    \mathrm{FT}\bigg[\frac{\partial^4}{\partial s_{1\parallel }^2\partial s_{2\parallel}\partial s_{3\parallel}} \langle w_{\parallel 1} w_{\parallel 2} \rangle \langle w_{\parallel 1} w_{\parallel 3} \rangle  +\mathrm{cyc.} \bigg] = \; &- \nu_3 k_3 \bigg[ \frac{\nu _1^3}{k_1}  \nu _2^4 P_{\theta\theta}(k_1) P_{\theta\theta}(k_2) + (1 \leftrightarrow 2) \bigg] \\ &+ \mathrm{cyc.}
    \end{split}
  \end{equation}
\end{itemize}
Here, it is understood that also the right-hand side of the expressions is evaluated up to fourth order in the linear perturbations. To summarise, the main difference between the large-scale limit of the streaming model and SPT is that the former approximates the transformation from real- to redshift-space without making extra assumptions about how the density and velocity fields evolve over time. To recover the SPT bispectrum it is therefore necessary to apply an additional small-amplitude expansion to the perturbations.  Importantly, when comparing the large-scale limit of the streaming model to SPT at any fixed order, $n_{\rm SPT}$, this small-amplitude expansion discards terms of the form $\langle w_{i\parallel}^2 \rangle\,\partial^2 \zeta_{\rm s}^{(n_{\rm SPT})}/\partial s_{i\parallel}^2$.  Such terms are not strictly subdominant since the zero-lag dispersions $\langle w_{i\parallel}^2\rangle$ receive contributions not only from large-scale bulk flows, but also from virialised velocities within collapsed structures, as noted, e.g., in \cite{scoccimarro2004,reid2011} (see also Section~\ref{sec: moments}).  These dispersions blur structures along the LOS in redshift space, leading to a damping of the large-scale power spectrum or bispectrum.  This is known as the Fingers-of-God (FoG) effect and will be further discussed in Section~\ref{sec:FoG}.

\subsection{Large-scale limit of the GSM and LSM}
\label{sec:LL_GSM_LSM}

Let us now discuss the large-scale limit under the assumption of specific velocity distributions for pairs and triplets, along with the LO expressions for their moments. This differs from the large-scale limit derived in the previous section, because only certain moments contribute to Eq.~(\ref{eq: lsl general}), and the LO approximation is applied to the moments $\mu_{m_1 m_2 m_3}^{(3)}$ and $\mu_{m_1 m_2}^{(2)}$ directly, instead of the full expression within the curly brackets of Eq.~(\ref{eq: lsl general}).  This makes it interesting to compare with the SPT result, and to do so, we again keep terms up to the fourth order in the perturbations and Fourier transform to the bispectrum.  We then obtain the difference between SPT and the Gaussian streaming model with LO moments, denoted as $\Delta B_{\rm s}^{\rm GSM} = B_{\rm s}^{\rm SPT} - B_{\rm s}^{\rm GSM}$:
\begin{align}
  \label{eq: LSL GSM FT}
  \Delta B_{\rm s}^{\rm GSM}(k_1,k_2,k_3) \simeq & \; \Big\{ \nu_1^2\,B_{\theta\delta\delta}(k_1,k_2,k_3) + \nu_1^2\,\nu_2^2\,B_{\theta\theta\delta}(k_1,k_2,k_3) - \nu_1^2\,\nu_2^2 \, P_{\delta\theta}(k_1)\,P_{\delta\theta}(k_2) \Big. \nonumber \\
                                                 & \Big. \hspace{0.8em} - \left[ \nu_1^2\,P_{\delta\theta}(k_1)\,P_{\delta\delta}(k_2) + \left(1 \leftrightarrow 2\right)\right] + \mathrm{cyc.} \Big\} + \nu_1^2\,\nu_2^2\,\nu_3^2\,B_{\theta\theta\theta}(k_1,k_2,k_3)\,.
\end{align}
We see that the large-scale limit of the GSM includes additional terms proportional to $P_{\delta\theta}^2$ and $P_{\delta\theta}\,P_{\delta\delta}$, but lacks all three bispectrum contributions.  Among these, $B_{\theta\theta\theta}$ is the only one directly related to the skewness of the triplewise velocity PDF.   The other differences arise from NLO contributions to either the mean LOS velocity, or its dispersion (cf. Eqs.~\ref{eq: lsl n=1} and \ref{eq: lsl n=2}).  

\begin{figure}[t]
  \centering
  \includegraphics[width=1.\textwidth]{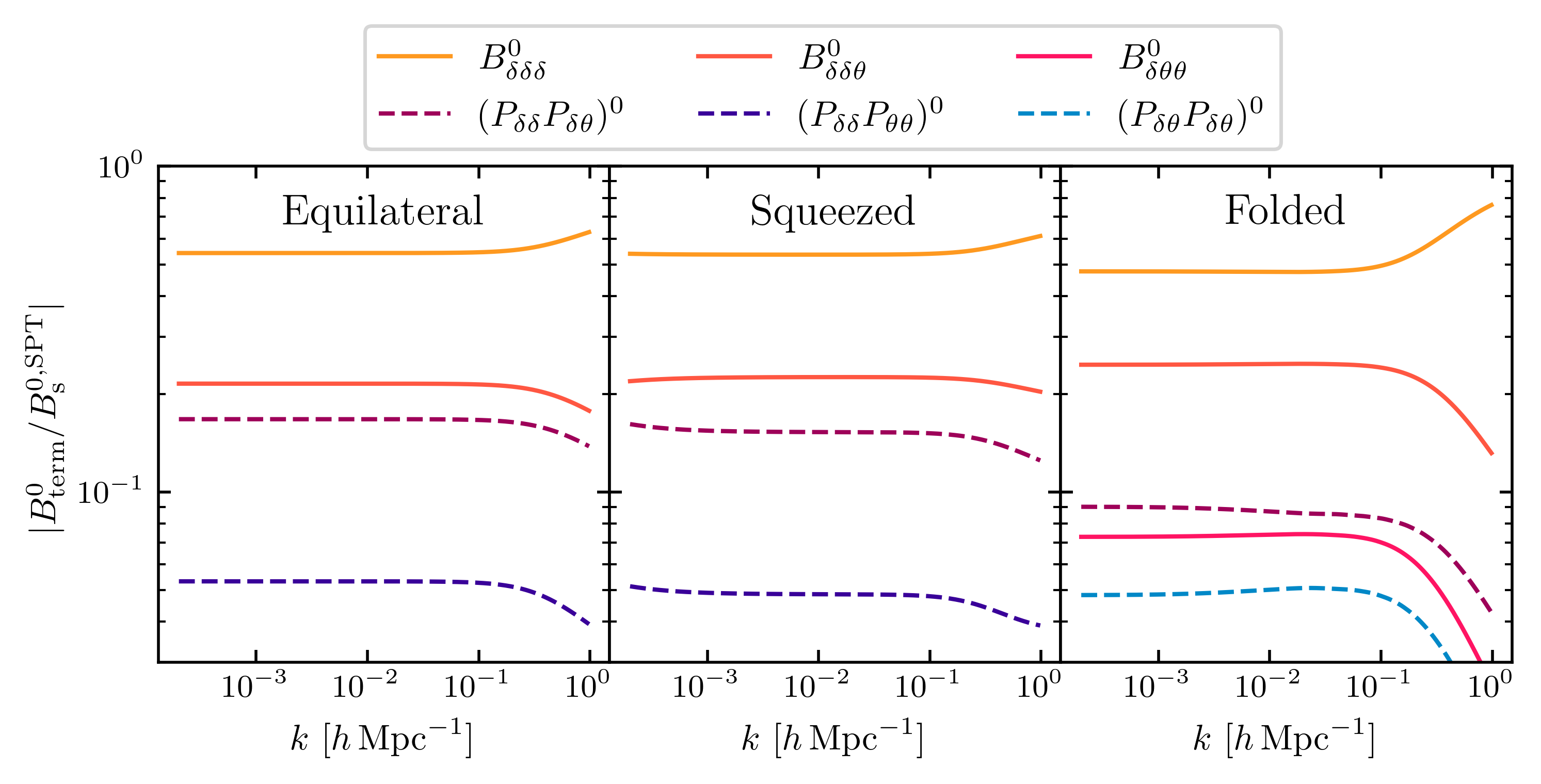}
  \caption{Fractional contributions of the six most relevant terms from Eq.~(\ref{eq: SPT bispectrum}) to the redshift-space bispectrum monopole in SPT. Solid lines represent contributions proportional to $B_{\alpha\beta\gamma}(k_1, k_2, k_3)$, dashed lines power spectra contributions proportional to $P_{\alpha\beta}(k_1)P_{\gamma\epsilon}(k_2)$.   \textit{Left:} equilateral configuration $(k, k, k)$. \textit{Centre:} squeezed configuration $(k, k, \delta k \ll k)$. \textit{Right:} folded configuration $(k, k, 2 k)$.  }
  \label{Fig: SPT contributions}
\end{figure} 
In order to understand the implications of the mismatch between SPT and the GSM, we compare the relative size of the various contributions to the total redshift-space bispectrum in SPT, see Eq.~(\ref{eq: SPT bispectrum}).
This is shown in Figure \ref{Fig: SPT contributions} for three different triangle shapes, where, for visualisation purposes, the terms have been averaged over all possible triangle orientations with respect to the LOS.  
In all cases, the main contribution beyond the real-space bispectrum, $B_{\delta\delta\delta}$, is the cross bispectrum between densities and velocities, $B_{\delta\delta\theta}$, followed by the double power spectrum term $P_{\delta\delta} P_{\delta\theta}$. All remaining terms have a smaller impact which varies depending on triangle shape. In particular, Figure ~\ref{Fig: SPT contributions} demonstrates that the impact of $B_{\theta\theta\theta}$ on the redshift-space bispectrum is minor, suggesting that assuming Gaussian PDFs does not significantly compromise the model's accuracy in the large-scale limit. However, the other terms, in particular $B_{\delta\delta\theta}$, contribute more significantly, which explains why the GSM with LO moments does not accurately match the redshift-space 3PCF and performs worse than SPT on large scales.  When instead applying the GSM and LO moments assumption to the 2PCF, the large-scale limit is consistent with SPT \cite{scoccimarro2004, cuesta2020} and therefore produces more accurate results, as we have seen in Section~\ref{sec: results}.

Not surprisingly, the situation for the Laplace streaming model is very similar.  The difference between the two in Fourier space is given by
\begin{align}
 B_{\rm s}^{\rm LSM}(k_1,k_2,k_3) - B_{\rm s}^{\rm GSM}(k_1,k_2,k_3) =  - \nu_3 k_3 \bigg[ \frac{\nu _1^3}{k_1}  \nu _2^4 P_{\theta\theta}(k_1) P_{\theta\theta}(k_2) + (1 \leftrightarrow 2) \bigg] + \mathrm{cyc.} \,,
\end{align}
and originates from the non-zero (excess) kurtosis of the Laplace distribution. Like the skewness term $B_{\theta\theta\theta}$ above, this term is very suppressed compared to the other contributions (and does not appear in Fig.~\ref{Fig: SPT contributions}), which is why in the large-scale limit the Gaussian and Laplace streaming models are almost identical (see Section~\ref{sec: gsm}).

\subsection{Connection between the streaming model and FoG damping functions}
\label{sec:FoG}

What remains to be understood from our results in Section~\ref{sec: results} is why the GSM with LO velocity moments is more accurate than SPT on smaller scales and especially for nearly isosceles configurations of the 3PCF multipoles.  As we show next, this increased accuracy is due to the GSM's non-perturbative treatment of the real- to redshift-space mapping, which together with the finite large-scale limit of the pair- and triplewise velocity dispersions results in a noticeable FoG effect across all scales.  As seen in Section~\ref{sec:LL_SPT}, SPT systematically neglects terms that are crucial to accurately describe this effect.  At order $n_{\rm SPT}$, these terms are of the form $\langle w_{i\parallel}^2 \rangle\,\partial^2 \zeta_{\rm s}^{(n_{\rm SPT})}/\partial s_{i\parallel}^2$, which, in Fourier space, become $-(k_i\,\nu_i)^2\,\sigma_v^2\,B_{\rm s}^{(n_{\rm SPT})}$, with $\langle w_{i\parallel}^2 \rangle = \sigma_v^2$.  Using a value $\sigma_v^2 \approx 15\,\left(h^{-1}\mathrm{Mpc}\right)^2$, which is consistent with our measurements presented in Section~\ref{sec: moments}, we estimate that the neglected terms give corrections of the order $10\,\%$ or larger for $k_i\,\nu_i \gtrsim 0.08\,h\,\mathrm{Mpc}^{-1}$.  Consequently, they should be accounted for already on quasi-linear scales for accurate model predictions.

In order to improve the SPT predictions one could retain a subset of the dispersion terms from higher expansion orders. This is why EFT approaches introduce additional redshift-space \emph{counterterms}, as proposed for the bispectrum in \cite{Ivanov2022}. However, the exact motivation for their counterterm prescription is not entirely clear, and it is worth noting that the scale-dependence of their counterterms differs from that of the dispersion terms identified above.

Alternatively, it is also possible to resum the effect of the dispersion terms to all orders in SPT under the assumption of specific pair- and triplewise velocity distributions, which gives rise to approximate FoG forms that are often used in the literature.  For instance, in case of the GSM we can express the full PDF as the convolution of two Gaussians,
  \begin{equation}
    \label{eq:PDFconv}
    {\cal P}^{(3)}(w_{12\parallel},w_{23\parallel}|\triangler) = \int {\cal P}^{(3)}_{\sigma_v}(u_{12\parallel},u_{23\parallel}) \, {\cal P}^{(3)}_{\text{no-}\sigma_v}(w_{12\parallel}-u_{12\parallel},w_{23\parallel}-u_{23\parallel}|\triangle_r) \, \mathrm{d}u_{12\parallel}\,\mathrm{d}u_{23\parallel}\,,
  \end{equation}
  whose individual means and covariances combine linearly, such that $\langle w_{12\parallel}\rangle_\triangle = \langle w_{12\parallel}\rangle_{\triangle,\sigma_v} + \langle w_{12\parallel}\rangle_{\triangle,\text{no-}\sigma_v}$ and $\boldsymbol{\mathsf{C}} = \boldsymbol{\mathsf{C}}_{\sigma_v} + \boldsymbol{\mathsf{C}}_{\text{no-}\sigma_v}$ (and analogously for the pairwise velocity distribution).  Let us now define the first Gaussian to have zero mean, $\langle w_{12\parallel}\rangle_{\triangle,\sigma_v} = 0$, and a covariance matrix that matches the (constant) large-scale limit of the full covariance, i.e.,
  \begin{equation}
    \label{eq:covLL}
    \boldsymbol{\mathsf{C}}_{\sigma_v} \equiv 2\sigma_v^2\,\left(
      \begin{array}{cc}
        1 & -1/2 \\
        -1/2 & 1
      \end{array}
    \right)\,.
  \end{equation}
  The second PDF, ${\cal P}^{(3)}_{\text{no-}\sigma_v}$, has accordingly the same mean as the full PDF, but a covariance that vanishes in the large-scale limit, meaning that it does not generate the terms that SPT fails to capture.  Plugging Eq.~(\ref{eq:PDFconv}) into the tree-point streaming model (Eq.~\ref{eq:3pcf sm}) we have:
  \begin{align}
    \label{eq:zeta_sv_conv}
    \zeta_{\rm s}(\triangles)
    &= \int {\cal P}^{(3)}_{\sigma_v}(u_{12\parallel},u_{23\parallel}) \, \Bigg\{ \int {\cal P}^{(3)}_{\text{no-}\sigma_v}(w_{12\parallel}-u_{12\parallel},w_{23\parallel}-u_{23\parallel}|\triangle_r) \Bigg. \nonumber \\
    & \hspace{2.5em} \Bigg. \times \, \Big[1+\xi(r_{12}) +\xi(r_{23}) +\xi(r_{31})+\zeta(\triangler) \Big]  \, \mathrm{d}w_{12\parallel} \, \mathrm{d}w_{23\parallel} \Bigg\}  \, \mathrm{d}u_{12\parallel}\,\mathrm{d}u_{23\parallel} \nonumber \\
    & \hspace{1.5em} - \Bigg[\int {\cal P}^{(2)}_{\sigma_v}(u_{12\parallel}) \Bigg\{ \int {\cal P}^{(2)}_{\text{no-}\sigma_v}(w_{12\parallel}-u_{12\parallel}|\bfr_{12})\,\Big[1 + \xi(r_{12})\Big] \, \mathrm{d}w_{12\parallel}\Bigg\} \, \mathrm{d}u_{12\parallel} + \mathrm{cyc.} \Bigg] - 1 \nonumber \\
    &= \int {\cal P}^{(3)}_{\sigma_v}(u_{12\parallel},u_{23\parallel}) \, \Bigg\{ \int {\cal P}^{(3)}_{\text{no-}\sigma_v}(w_{12\parallel}-u_{12\parallel},w_{23\parallel}-u_{23\parallel}|\triangle_r) \Bigg. \nonumber \\
    & \hspace{2.5em} \Bigg. \times \, \Big[1+\xi(r_{12}) +\xi(r_{23}) +\xi(r_{31})+\zeta(\triangler) \Big]  \, \mathrm{d}w_{12\parallel} \, \mathrm{d}w_{23\parallel} \nonumber \\
    & \hspace{1.5em} \Bigg. - \Bigg[ \int {\cal P}^{(2)}_{\text{no-}\sigma_v}(w_{12\parallel}-u_{12\parallel}|\bfr_{12})\,\Big[1 + \xi(r_{12})\Big] \, \mathrm{d}w_{12\parallel} + \mathrm{cyc.} \Bigg] -1 \Bigg\} \, \mathrm{d}u_{12\parallel}\,\mathrm{d}u_{23\parallel} \nonumber \\
    &= \int {\cal P}^{(3)}_{\sigma_v}(u_{12\parallel},u_{23\parallel}) \, \zeta_{\text{s,no-}\sigma_v}(\triangles - \triangle_{u_\parallel}) \, \mathrm{d}u_{12\parallel}\,\mathrm{d}u_{23\parallel}\,,
  \end{align}
  where $\triangles - \triangle_{u_{\parallel}}$ denotes the shift of the positions $\bfs_{12}$, $\bfs_{23}$, and $\bfs_{31}$ due to the LOS components $u_{12\parallel}$ and $u_{23\parallel}$.  We have repeatedly used that ${\cal P}^{(3)}_{\sigma_v}$ is scale-independent (and thus can be separated from the integrations over $w_{12\parallel}$ and $w_{23\parallel}$) and between the first and second equal sign we have, moreover, exploited that a marginalised multivariate Gaussian is again a Gaussian\footnote{This property holds for any multivariate PDF whose characteristic function depends on the combination $(u_{12\parallel},u_{23\parallel})^{\rm T} \cdot \boldsymbol{\mathsf{C}}_{\sigma_v}^{-1} \cdot (u_{12\parallel},u_{23\parallel})$.}, i.e., $\int {\cal P}^{(3)}_{\sigma_v}(u_{12\parallel},u_{23\parallel})\,\mathrm{d}u_{23\parallel} = {\cal P}^{(2)}_{\sigma_v}(u_{12\parallel})$, as well as that\footnote{This is true only for the specific correlation structure of $\boldsymbol{\mathsf{C}}_{\sigma_v}$.} ${\cal P}^{(3)}_{\sigma_v}(u_{12\parallel},-u_{12\parallel}-u_{23\parallel}) = {\cal P}^{(3)}_{\sigma{_v}}(u_{12\parallel},u_{23\parallel})$.  Assuming the GSM, we can represent the complete redshift-space 3PCF as the convolution of a 3PCF, $\zeta_{\text{s,no-}\sigma_v}$, which perturbation theory can model in the large-scale limit without omitting important terms, together with a PDF for the one-point velocity dispersions.

In Fourier space, the convolution in Eq.~(\ref{eq:zeta_sv_conv}) becomes a product, with the Fourier transform of the one-point velocity dispersion distribution being
\begin{equation}
  \label{eq:GaussDamping}
    \mathrm{FT}\left[{\cal P}^{(3)}_{\sigma_v}\right](k_{1\parallel},k_{2\parallel},k_{3\parallel}) 
    = \exp{\left[-\frac{\sigma_v^2}{2}\left(k_{1\parallel}^2+k_{2\parallel}^2+k_{3\parallel}^2\right)\right]} \equiv D_{\rm FoG}(\bfk_1, \bfk_2, \bfk_3)\,,
  \end{equation}
 \textit{motivating} a model for the FoG-damped bispectrum, as proposed e.g. in \cite{ScoCouFri9906, HasRasTar1708}:
  \begin{equation}
    B_{\rm s}(\bfk_1, \bfk_2, \bfk_3) =   D_{\mathrm{FoG}}(\bfk_1, \bfk_2, \bfk_3) \,  B_{\rm s}^{\rm SPT}(\bfk_1, \bfk_2, \bfk_3)\,.
  \end{equation}
  We need to note, however, that $B_{\rm s}^{\rm SPT}$ is not exactly the Fourier transform of $\zeta_{\text{s,no-}\sigma_v}$ since it does contain terms relating to skewness, kurtosis, etc., which are missing in the GSM.  Furthermore, beyond tree-level order terms relating to the \textit{linear} velocity dispersion should be subtracted from $B_{\rm s}^{\rm SPT}$ (as done in \cite{HasRasTar1708}) to avoid double-counting with the terms included through the damping factor.

Different assumptions about the pair- and triplewise PDFs lead to different resummations of the dispersion terms and accordingly to different FoG damping functions.  Adopting the Laplace streaming model for ${\cal P}^{(3)}$ as in Eq.~(\ref{eq:PDF_L}), and also a Laplace distribution\footnote{In this case the distribution for ${\cal P}_{\text{no-}\sigma_v}^{(3)}$ is not Laplace, though.} for ${\cal P}_{\sigma_v}^{(3)}$, we obtain a Lorentzian damping function
\begin{align}
  D_{\rm FoG}(\bfk_1, \bfk_2, \bfk_3) = \frac{1}{1+\sigma_v^2\,\left(k_{1\parallel}^2+k_{2\parallel}^2+k_{3\parallel}^2\right)/2}\,.
\end{align}
In the large-scale limit (quadratic order in the wavenumbers), the Gaussian and Lorentzian damping functions are identical and they only start to differ by $\gtrsim 10\,\%$ on scales larger than $\bar{k} \sim 0.3\,h\,\mathrm{Mpc}^{-1}$, with $\bar{k} = \sqrt{k_{1\parallel}^2+k_{2\parallel}^2+k_{3\parallel}^2}$, or alternatively, on scales $\lesssim 10\,h^{-1}\,\mathrm{Mpc}$.  This is consistent with our numerical result from Section~\ref{sec: gsm} and suggests that the precise functional form of the triplewise PDF becomes relevant on scales much smaller than the overall damping (FoG) effect.

\begin{figure}[t]
\centering
\includegraphics[width=1.\textwidth]{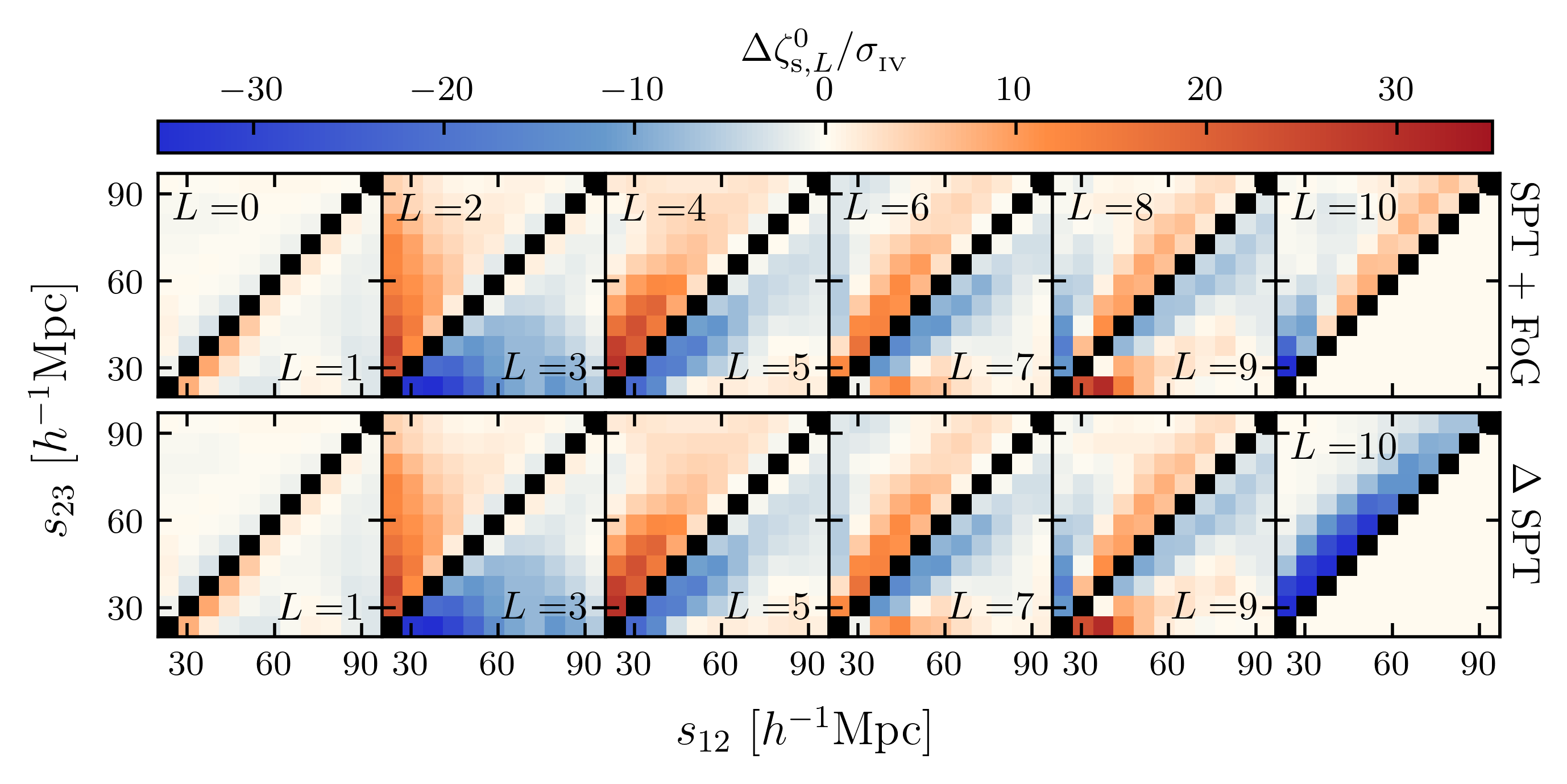}
    \caption{ Differences between measured 3PCF multipoles and the SPT model with Gaussian damping function (top), and between SPT with and without the damping function (bottom).  In both cases, the differences are shown in units of standard deviations corresponding to a Stage IV-like volume.
    }
    \label{Fig: spt FoG diff multipoles}
\end{figure}

Finally, Figure~\ref{Fig: spt FoG diff multipoles} illustrates the impact of resumming the finite dispersion terms in the SPT model using a Gaussian damping function (Eq.~\ref{eq:GaussDamping}).  The agreement with the measured 3PCF multipoles (top row) is significantly improved across all multipoles compared to the SPT model without damping, as shown in Figure~\ref{Fig: model comparison spt gsm mm diff multipoles}.  Especially for small-scale and nearly isosceles configurations, the performance is now similar to the GSM with LO SPT moments.  This is precisely where the FoG effect is most significant, as indicated by the bottom row, which plots the difference between the SPT with and without the damping function.  The effect is concentrated on small scales for the lower multipoles, whereas it extends to larger scales, but only for the nearly isosceles configurations for the higher multipoles.  This is due to a combination of two factors: 1) as noted before, isosceles and nearly isosceles configurations of the 3PCF multipoles contain triangles whose third side ranges from $|s_{12} - s_{23}|$ to $s_{12} + s_{23}$, making them subject to highly non-linear scales; 2) higher-order multipoles measure faster variations (higher powers) of $\hat{\bfs}_{12} \cdot \hat{\bfs}_{23}$, which are generated by the non-perturbative damping function on increasingly non-linear scales.

\section{Discussion and conclusions}
\label{sec: conclusion}

In this work, we conducted a detailed analysis of the matter three-point correlation function (3PCF) in redshift space, focusing on two models that employ different assumptions for the mapping from real to redshift space.  Specifically, we examined:
  \begin{enumerate}
  \item The streaming model, which accounts for the fully non-linear mapping but requires assuming a functional form for the PDF of the pair- and triplewise velocity distributions. In particular,
  we consider the Gaussian approximation which leads to the Gaussian streaming model (GSM).
  \item Standard perturbation theory (SPT), which expands the non-linear mapping perturbatively, assuming sufficiently small density and velocity fluctuations.
  \end{enumerate}
Our objective was to compare the outcomes of these models against N-body simulations and also determine the regimes in which they either produce similar or differing predictions.  To achieve this, we  extracted the monopole of the redshift-space 3PCF from the Quijote simulations, as well as the first two moments of the pair- and triplewise velocity distributions.  This allowed us to directly analyse the modelling of the 3PCF, but also of the two main ``ingredients'' of the GSM and LSM.  Furthermore, for the first time, we evaluated the three-point streaming model in a well-defined large-scale limit,  demonstrating the conditions under which it aligns with SPT.  Our findings can be summarised as follows.

\begin{itemize}
\item Assuming a Gaussian PDF for the triplewise velocity distribution does not significantly impact the monopole of the 3PCF on scales larger than approximately $10\,h^{-1}\,\mathrm{Mpc}$.   Although the actual velocity distribution can deviate considerably from a Gaussian \cite{kuruvilla2020}, comparing the GSM with a streaming model that uses a Laplace triplewise velocity distribution (which has heavily enhanced tails) shows no noticeable differences on these scales (see Figure~\ref{Fig: LSM vs GSM}).  This is because, in this regime, the large-scale limit applies, where the line of sight displacement due to the redshift-space mapping is small compared to the separation between the objects.  We have shown that in the large-scale limit, the 3PCF is dominated by the first two velocity moments: the mean infall velocity and its dispersion.  Therefore, accurately modelling these moments can provide an accurate prediction of the 3PCF, regardless of the shape of the PDF.
\item Unlike the 2PCF, which is directly sensitive to the infall velocity and dispersion, the 3PCF measures how these pairwise moments are affected by the presence of the third object in the triplet.  We have demonstrated this modulation in all of the independent components of the infall velocity and dispersion tensor, identifying several components for the latter that had been overlooked in the previous work by \cite{kuruvilla2020} as they vanish at leading order (LO) in SPT.
When comparing our measurements with LO SPT predictions (second order in the perturbations), we find good agreement on scales $\gtrsim 50\,h^{-1}\,\mathrm{Mpc}$ (see~\Cref{Fig: Moments pairwise,Fig: Moments trip first,Fig: Moments trip second,Fig: Moments trip second mixed}), after correcting the predictions for the velocity dispersion with a non-perturbative additive constant.  However, the modulation is not fully captured because it is partially linked to cross-bispectra between densities and velocities, which vanish when restricted to second-order perturbations.
\item Due to this insufficient modelling, the GSM combined with LO SPT expressions for the velocity moments performs worse for the 3PCF than an equivalent model for the 2PCF.  Specifically, we observe significant discrepancies relative to measurement uncertainties for a Stage-IV survey, even on relatively large scales ($\sim 60\,h^{-1}\,\mathrm{Mpc}$, see~\Cref{Fig: model comparison jb gsm ptm,Fig: model comparison spt gsm mm diff multipoles}).  We confirmed that these issues are indeed due to the modelling of the moments by replacing the LO expressions with the measured moments, which leads to much more accurate results, especially at small scales and for scalene configurations.
\item The model based entirely on tree-level SPT improves on the GSM with LO moments on scales larger than $\sim 60\,h^{-1}\,\mathrm{Mpc}$, but is significantly worse on smaller scales and for (nearly) isosceles configurations (see~\Cref{Fig: model comparison jb gsm ptm,Fig: model comparison spt gsm mm diff multipoles}).  Consequently, there is no regime where these two models agree with each other.  This discrepancy arises because SPT includes the aforementioned cross-bispectra, allowing it to more accurately model the modulation of the velocity moments.  However, the deficiency observed on smaller scales and for isosceles configurations is due to the perturbative treatment of the redshift-space mapping in SPT (see~\Cref{Fig: spt FoG diff multipoles}).  In particular, the finite large-scale limit of the velocity dispersion gives rise to terms that SPT systematically neglects but which are naturally included in the GSM formulation.  These terms are responsible for the Fingers-of-God (FoG) effect and by separating them in either the Gaussian or Laplace streaming model, we showed how to motivate Gaussian or Lorentzian FoG damping functions commonly used in the literature.
\end{itemize}

To conclude, the combination of the tree-level SPT model with a FoG damping function leverages the strengths of the two theoretical approaches studied in this work.  Among these, it appears to be the preferred choice for modelling the redshift-space 3PCF.  When evaluating the 3PCF decomposed into multipoles with respect to the angle between $\bfs_{12}$ and $\bfs_{23}$ 
(which is necessary to match the estimates of the 3PCF from the simulations produced by the numerically efficient \textsc{Encore} code), this model is also computationally advantageous.  The multipoles can be efficiently computed from the Fourier space bispectrum using the 2D-FFTLog algorithm, which we validated in detail as part of this work (Appendix~\ref{app: 2dfftlog}). In contrast, evaluating the streaming model 
can be computationally rather demanding, which would limit its use in Bayesian inference applications.

Our work also highlights that there is further room for improvement in the GSM by focusing efforts on a better modelling of the triplewise infall velocity and dispersion.  Moreover, it will be interesting to study the quadrupole (with respect to the line of sight) of the 3PCF.  The quadrupole is a more sensitive probe of redshift-space distortions and, as such, might show a stronger dependence on the shape of the triplewise velocity PDF than what we found for the monopole in this work.  We reserve both of these avenues for future research.

\acknowledgments
AP is a part of the International Max Planck Research School in Astronomy and Astrophysics, the Bonn Cologne Graduate School, and a guest at the Max Planck Institute for Radio Astronomy in Bonn.
AE is supported at the AIfA by an Argelander Fellowship.
CP is grateful to SISSA, the University of Trieste, and IFPU, where part of this work was carried out, for hospitality and support.
In addition to the software cited in the main text, this research made use of: \textsc{NumPy} \cite{Harris2020}, \textsc{matplotlib} \cite{Hunter2007}, \textsc{SciPy} \cite{Virtanen2020}, \textsc{GNU scientific library} \cite{gough2009}, \textsc{h5py} \cite{hdf52014}, \textsc{mpi4py} \cite{Dalcin2021}.

\bibliographystyle{bibstyle_jcap}
\bibliography{main}

\appendix
\section{SPT kernels}

For reference, we provide here the redshift-space kernel functions, $Z_n(\bfk_1, \dots , \bfk_n)$, up to order $n=2$, which are required in the computation of the SPT bispectrum, see Eq.~(\ref{eq: spt bispectrum kernels}).  They take the following form \cite[e.g.,][]{bernardeau2002}:
\begin{align}
Z_1(\bfk_1) &= 1+ f \mu_1^2  \,, \label{eq:Z1} \\
Z_2(\bfk_1, \bfk_2) &= F_2(\bfk_1,\bfk_2) + f \mu^2 G_2(\bfk_1,\bfk_2)+ f\frac{\mu k }{2}\bigg[ \frac{\mu_1}{k_1}(1+f\mu_2^2)+ \frac{\mu_2}{k_2}(1+f\mu_1^2)\bigg]\,, \label{eq:Z2}
\end{align}
where $\mu_i= \bfk_i \cdot \zhat$,  $\bfk = \bfk_1 + \bfk_2 = - \bfk_3$ and $\mu = \bfk \cdot \zhat/k = - \mu_3 $, with $\zhat$ denoting the LOS.  The real-space second-order kernels for the density and velocity perturbations are given by:
\begin{align}
\label{eq:real space kernels}
       F_2(\bfk_1,\bfk_2) & =\frac{5}{7}+ \frac{1}{2}\frac{\bfk_1\cdot \bfk_2}{k_1 k_2} \bigg( \frac{k_1}{k_2}+ \frac{k_2}{k_1} \bigg) +\frac{2}{7} \bigg( \frac{\bfk_1 \cdot \bfk_2}{k_1 k_2} \bigg)^2  \,, \\
       G_2(\bfk_1,\bfk_2) &=\frac{3}{7}+ \frac{1}{2}\frac{\bfk_1\cdot \bfk_2}{k_1 k_2} \bigg( \frac{k_1}{k_2}+ \frac{k_2}{k_1} \bigg) +\frac{4}{7} \bigg( \frac{\bfk_1 \cdot \bfk_2}{k_1 k_2} \bigg)^2 \,.
\end{align}
For any further details on SPT, we refer the reader to the review \cite{bernardeau2002}.

\section{Real-space tree-level 3PCF}
\label{sec:app.real-space-tree}

The evaluation of the streaming model (see Section~\ref{sec: gsm}) requires a model for the 3PCF in real space.  For this purpose and throughout all this work, we have used the tree-level, infrared resummed SPT prediction, which is obtained by Fourier transforming the corresponding bispectrum
\begin{equation}
\label{eq: tree level real space B}
    B(\bfk_1,\bfk_2,\bfk_3)=2 F_2(\bfk_1,\bfk_2) \plir(k_1) \plir(k_2) + \mathrm{cyc.}\,,
\end{equation}
where $\plir$ is the linear, infrared resummed power spectrum as defined in Section~\ref{sec:IRresum}.  It was shown in \cite{jing1996} that the Fourier transform of the tree-level SPT bispectrum can be written in terms of the 2PCF, as well as two further one-dimensional quantities, $\eta_l(r)$ and $\epsilon_l(r)$, which are given by:
\begin{eqnarray}
  \label{eq: eta}
  \eta_l(r) &=& \frac{1}{2 \pi^2} \int  k^2 \frac{kr \cos{(kr)} - \sin{(kr)}}{kr^3}  \frac{\plir(k)}{k^l}\, \mathrm{d}k \,, \\
  \label{eq: epsilon}
  \epsilon_l(r) &=& \frac{1}{2\pi^2} \int k^2 \frac{3(\sin{(kr)} - kr \cos{(kr)}) - k^2 r^2 \sin{(kr)}}{kr^5} \frac{\plir(k)}{k^l} \, \mathrm{d}k \,.
\end{eqnarray}
With these definitions, we finally arrive at the following expression for the real-space 3PCF:
\begin{eqnarray}
\label{eq: jb}
    \zeta(\triangler) &=& \bigg\{ \frac{10}{7} \, \xi(r_{12})\,\xi(r_{23}) + \big[ \eta_2(r_{12})\,\eta_0(r_{31})  + \eta_0(r_{12}) \, \eta_2(r_{31})\big] \bfr_{12} \cdot \bfr_{31}  \nonumber\\
    && + \frac{4}{7} \big[ \epsilon_2 (r_{12})\,\epsilon_2 (r_{31}) (\bfr_{12} \cdot \bfr_{31} )^2 + \epsilon_2 (r_{12 }) \eta_2(r_{31}) r_{12}^2 + \eta_2(r_{12}) \epsilon_2 (r_{31 }) r_{31}^2 \nonumber\\
    &&+ 3  \eta_2(r_{12})\,\eta_2(r_{31})  \big] \bigg\} + \mathrm{cyc.}
\end{eqnarray}

\section{Pairwise and triplewise velocity moments in SPT}
\label{app: velocity moments}
In this subsection we summarise the leading-order (LO) SPT predictions for all independent components of the first two velocity moments, which we compare to our measurements in Section \ref{sec: moments}.

In the pairwise case, the mean velocity only has a component parallel to the separation vector $\bfr_{12}$ (denoted by the subscript ``r'', standing for radial), which at LO is given by:
\bea
 \label{eq: mean pair-wise spt}
     \langle w_{12 \mathrm{r}} | \bfr_{12} \rangle_\mathrm{p} 
    \lo    \frac{f}{\pi^2 }\int  \, k \, j_1(k r_{12}) \, \plir(k) \,\mathrm{d}k \equiv \bar{w}(r_{12}) \,.
\eea
Meanwhile, the pairwise dispersion contains both parallel and transverse components (the latter denoted by the subscript ``t''):
\bea
\label{eq: dispersion pair-wise spt}
    \langle w_{12 \mathrm{r}}^2 | \bfr_{12} \rangle_\mathrm{p} 
    &\lo &  2 [ \sigma_{v,\mathrm{lin}}^2 -  \psi_r(r_{12}) ] \;, \\
    \langle w_{12 \mathrm{t}}^2 | \bfr_{12} \rangle_\mathrm{p} 
    &\lo &  2 [ \sigma_{v,\mathrm{lin}}^2 -  \psi_p(r_{12}) ] \;,
    \eea
and the functions $\psi_r$ and $\psi_p$ are defined in Section~\ref{sec: gsm}.

In triplet configurations the third object adds a transverse component to the otherwise radial infall velocity between the other two objects.  At LO the two components are given by (see \cite{kuruvilla2020}):
\bea
 \label{eq: mean triple-wise spt}
     \langle w_{12 \mathrm{r}}  \rangle_\triangle
   & \lo&   \bar{w}(r_{12})-\frac{1}{2} \Big[\bar{w}(r_{23}) \cos{\chi} - \bar{w}(r_{31}) \frac{r_{12}+r_{23} \cos{\chi}}{r_{31}}\Big] \,, \\
    \langle w_{12 \mathrm{t}}  \rangle_\triangle  &\lo&   -\frac{1}{2}\Big[\bar{w}(r_{23}) - \bar{w}(r_{31})\frac{r_{23}}{r_{31}}\Big]\sin{\chi}\,,
\eea
where $\chi$ is the angle between $\bfr_{12}$ and $\bfr_{23}$. Analogue expressions hold for the velocity moments corresponding to the pair `23', with $r_{12} \leftrightarrow r_{23}$.
At LO in SPT, the radial and transverse components of the triplewise dispersions are identical to the pairwise case, while the component normal to the plane of the triangle (see discussion in Section~\ref{sec: moments}) equals the transverse one. Hence,
\bea
\label{eq: dispersion triple-wise spt}
    \langle w_{12 \mathrm{r}}^2 \rangle_\triangle
    &\lo &  2 [ \sigma_{v,\mathrm{lin}}^2 -  \psi_r(r_{12}) ] \;, \\
    \langle w_{12 \mathrm{t}}^2 \rangle_\triangle
    &\lo &  2 [ \sigma_{v,\mathrm{lin}}^2 -  \psi_p(r_{12}) ] \lo \langle w_{12\mathrm{n}}^2\rangle_{\triangle} \;.
\eea
The mixed radial-transverse component, $\langle w_{12\mathrm{r}}\,w_{12\mathrm{t}}\rangle$, vanishes at this order in the perturbations.  Finally, the five components of the mixed second-order moments are given by:
\bea
\label{eq: mixed triple-wise spt}
    \langle w_{12 \mathrm{r}} w_{23 \mathrm{r}} \rangle_\triangle
    &\lo & (\psi_r(r_{12})+\psi_r(r_{23})-\psi_p(r_{31})-\sigma_{v,\mathrm{lin}}^2) \cos{\chi} \nonumber\\
    &&- (\psi_r(r_{31})-\psi_p(r_{31})) \frac{r_{12}+r_{23}\cos{\chi}}{r_{31}} \frac{r_{12}\cos{\chi}+r_{23}}{r_{31}} \;, \\
    \langle w_{12 \mathrm{r}} w_{23 \mathrm{t}} \rangle_\triangle&\lo & -(\psi_r(r_{12})+\psi_p(r_{23})-\psi_p(r_{31})-\sigma_{v,\mathrm{lin}}^2)\sin{\chi} \nonumber\\
    &&+ (\psi_r(r_{31})-\psi_p(r_{31})) \frac{r_{12}+r_{23}\cos{\chi}}{r_{31}} \frac{r_{12} \sin{\chi}}{r_{31}} \;, \\
    \langle w_{12 \mathrm{n}} w_{23 \mathrm{n}} \rangle_\triangle&\lo & \psi_p(r_{12})+\psi_p(r_{23})-\psi_p(r_{31})-\sigma_{v,\mathrm{lin}}^2 \;, \\
    \langle w_{12 \mathrm{t}} w_{23 \mathrm{r}} \rangle_\triangle&\lo & (\psi_p(r_{12})+\psi_r(r_{23})-\psi_p(r_{31})-\sigma_{v,\mathrm{lin}}^2)\sin{\chi} \nonumber\\
    &&-(\psi_r(r_{31})-\psi_p(r_{31})) \frac{r_{23}\sin{\chi}}{r_{31}}\frac{r_{12}\cos{\chi}+r_{23}}{r_{31}} \;, \\
    \langle w_{12 \mathrm{t}} w_{23 \mathrm{t}} \rangle_\triangle&\lo & (\psi_p(r_{12})+\psi_p(r_{23})-\psi_p(r_{31})-\sigma_{v,\mathrm{lin}}^2)\cos{\chi} \nonumber\\
    &&+(\psi_r(r_{31})-\psi_p(r_{31})) \frac{r_{23} \sin{\chi}}{r_{31}}\frac{r_{12}\sin{\chi}}{r_{31}} \;.
\eea
The redshift-space mapping depends only on the LOS components of these velocity moments and the corresponding projections, as required by the evaluation of the streaming model, are given in Section~\ref{sec: gsm}.

\section{Convergence tests: Subsampling and size of random catalogue} 
\label{app:subsampling}

The measurement of the 3PCF from the full matter particle distribution is computationally very demanding, which is why we randomly subsampled the number of particles in the Quijote snapshots by a factor of $4^3$, corresponding to a total number of $N_\mathrm{p} = 128^3$ particles.  Moreover, 
as mentioned in Section \ref{sec: encore}, ENCORE measures the 3PCF by comparing the triplet counts in the data catalogue with those in a purely random distribution of particles.  We chose a random catalogue that is 10 times denser, i.e., containing 
$N_\mathrm{r}= 10 \times N_\mathrm{p}$ particles.
In this appendix we demonstrate that these choices do not affect the clustering properties and lead to converged measurements of the 3PCF.

\begin{figure}[t]
    \centering
\includegraphics[width=1.\textwidth]{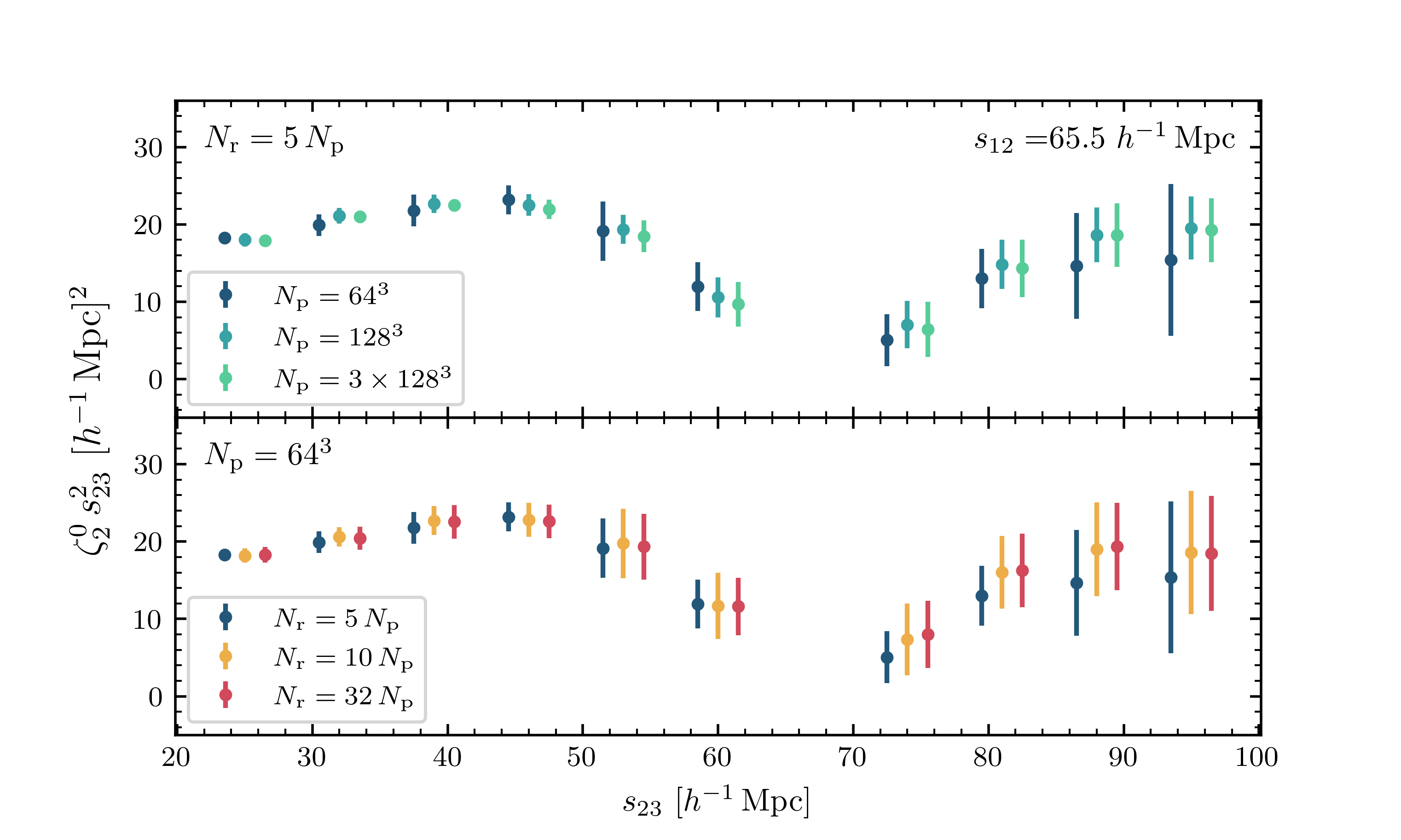}
    \caption{Quadrupole of the isotropic 3PCF estimated by ENCORE \cite{philcox2021encore} for a fixed bin of $s_{12} \in ( 62, 69) \, h^{-1} $ Mpc. We show the mean and standard deviation taken measuring the 3PCF from 30 realisations of the Quijote simulation. The points have been shifted along the x-axis to improve the readability of the plot. \textit{Top}: Impact of different subsamplings of the data $N_\mathrm{p}$ (original size : $N_\mathrm{p} = 512^3$). \textit{Bottom}: Impact of using different sizes, $N_\mathrm{r}$, for the random catalogue. }
    \label{Fig: data selection}
\end{figure}

We have tested four different subsamplings, ranging from $N_{\mathrm{p}} = 32^3$ to $3 \times 128^3$ particles.  While a subsampling to $N_\mathrm{p}=32^3$ particles yields clearly biased measurements, the remaining three cases are shown in the top panel of Figure~\ref{Fig: data selection}, which plots the quadrupole moment ($L=2$; with respect to the angle between $\bfs_{12}$ and $\bfs_{23}$) of the 3PCF monopole for a fixed bin $s_{12} \in ( 62, 69) \, h^{-1}\mathrm{Mpc}$.  The measurements have been repeated for 30 independent realisations and the plot shows the resulting mean and standard deviation.  We observe that between $N_{\rm p} = 64^3$ and $N_{\rm p} = 3 \times 128^3$ there are still significant differences, both in the means and the standard deviations, but these shrink to a level acceptable to our analysis when considering $N_{\rm p} = 128^3$ particles.  For a single realisation we have also tested a subsampling to $N_{\rm p} = 7 \times 10^6$ particles, which again led to results consistent with our nominal choice.
These results have been obtained using $N_\mathrm{r}=5 N_\mathrm{p}$ random points.

In the bottom panel of Figure~\ref{Fig: data selection} the number of subsampled particles has instead been fixed to $N_\mathrm{p}=64^3$ in order to test the impact of the size of the random catalogue.
Starting from $N_\mathrm{r}=5 N_\mathrm{p}$ random particles, we increase their number in two steps up to 32 times the particles contained in the data catalogue.  We find some disagreement in the measurements between the smallest and largest random catalogues considered (although still within the errorbars), whereas $N_\mathrm{r}=10 N_\mathrm{p}$ and $N_\mathrm{r}=32 N_\mathrm{p}$ give almost indistinguishable results. 

Although we have shown these tests exemplary only for the $L=2$ moment and for a particular configuration, we find very similar results for all other considered multipoles and configurations.  For that reason,
we work with a subsampling of $N_\mathrm{p}=128^3$ particles and $N_\mathrm{r}=10 \,N_\mathrm{p}$ random points for our analysis presented in the main text.

\section{Validation of the 2D-FFTLog method}
\label{app: 2dfftlog}

In this appendix we present the 2D-FFTLog algorithm and study how different parameter choices affect the computation of the 3PCF multipoles.

Computing the 3PCF multipoles, $\zeta_L$, from the analogue multipoles of the bispectrum, $B_L$, requires a two-dimensional integration over two Bessel functions $j_L$ (see Eq.~\ref{eq: multipoles FT}), i.e.
\begin{equation}
  \label{eq: FT 2d}
  \zeta_{L} (r_{12}, r_{23}) = (-1)^{L} 
  \int  \Delta_L (k_1,k_2)
  j_L(k_1 s_{12}) j_L(k_2 s_{23}) \, \frac{\mathrm{d} k_1}{k_1} \frac{\mathrm{d} k_2}{k_2} \,,
\end{equation}
where 
\begin{equation}
  \label{eq: Delta}
  \Delta_L (k_1,k_2) \equiv \frac{k_1^3 k_2^3}{(2 \pi^2)^2}  B_L (k_1, k_2) \,.
\end{equation}
Naive discretisation of this expression would take of the order of $N_r^2 \, N_k^2$ steps, where $N_r$ is the number of bins in either $r_{12}$ or $r_{23}$, and $N_k$ is the number of sampled $k$ values. Since the Bessel functions are rapidly oscillating, a very large number $N_k$ is typically required for the integral to converge. 
The 2D-FFTLog algorithm \cite{fang2021} has been proposed to overcome this issue and generalises the original FFT-Log approach by \cite{hamilton2000} to the two-dimensional integration of the kind of Eq.~(\ref{eq: FT 2d}).  It performs the integration in logarithmic $k$-bins, and decomposes the function $\Delta_L$ into power laws as follows:
\begin{equation}
  \label{eq: delta power law }
  \Delta_L (k_1,k_2) =  \frac{1}{N^2}  \sum\limits_{m=-N/2}^{N/2} \sum\limits_{n=-N/2}^{N/2} c_{L,mn} \bigg( \frac{k_1^{\nu_1+i \eta_m}}{k_{\rm min}^{i \eta_m}} \bigg) \bigg( \frac{k_2^{\nu_2+i \eta_n}}{k_{\rm min}^{i \eta_n}} \bigg) \,,
\end{equation}
where 
\begin{equation}
  \label{eq: cmn coeff}
  c_{mn} = \sum_{p=0}^{N-1} \sum_{q=0}^{N-1} \frac{\Delta_L (k_p, k_q)}{k_p^{\nu_1} k_q^{\nu_2} } e^{-2 \pi i ( m p + n q )/N}    \,,
\end{equation}
and 
\begin{equation}
  \label{eq: eta_m}
  \eta_m=2 \pi \frac{m}{N \Delta_{\ln{k}}}\,.
\end{equation}
The logarithmic $k$-bins are defined by $k_p = k_{\rm min}\,\exp{(p\,\Delta_{\ln{k}})}$, with $k_{\rm min}$ being the smallest value considered and $\Delta_{\ln{k}}$ the logarithmic bin size.  The parameters $\nu_1$ and $\nu_2$ are rational numbers and are called ``bias'' parameters. Taking the $r$- and $k$-arrays to have the same number of bins $N = N_r = N_k$, as well as the same logarithmic spacing, $ \Delta_{\ln{r}}=\Delta_{\ln{k}} $, we can rewrite Eq.~(\ref{eq: FT 2d}) in the form 

\begin{equation}
\label{eq: IFFT2}
    \zeta_{L} (r_{12}, r_{23}) = (-1)^{L} \frac{\pi}{16 r_{12}^{\nu_1} r_{23}^{\nu_2} } \mathrm{IFFT2}[c_{L,mn}^* ( k_{\rm min}\,r_{12})^{i \eta_m} ( k_{\rm min}\,r_{23})^{i \eta_n} g_L(\nu_1 -i \eta_m) g_L(\nu_2 -i \eta_n) ]
\end{equation}
where $\mathrm{IFFT2}$ stands for the two-dimensional inverse Fast Fourier Transform, and the functions $g_L$ are defined as 
\begin{equation}
\label{eq: g_l}
    g_L(\omega)= 2^\omega \frac{\Gamma(\frac{L+\omega}{2})}{\Gamma(\frac{3+L-\omega}{2})}, \hspace{1cm} -L < \mathcal{R}(\omega) < 2 \,. 
\end{equation}
The valid ranges of the bias parameters are therefore $-L_1 < \nu_1 < 2$, $-L_2 < \nu_2 < 2$.  Different choices of bias parameter values can cause different levels of ringing effects at the edges of the $r$-array \cite{fang2021}, as we consider below. 

To study the accuracy of the 2D-FFTLog algorithm and to avoid ringing effects we investigate the impact of the different parameters entering in the expressions above. We vary each of them at a time and compare the result against the exact reference for the real-space 3PCF from Appendix~\ref{sec:app.real-space-tree}.  In the latter case we use Gauss-Legendre integration to compute the multipoles from the full 3PCF.
  
\begin{figure}[t]
    \centering
\includegraphics[width=1.\textwidth]{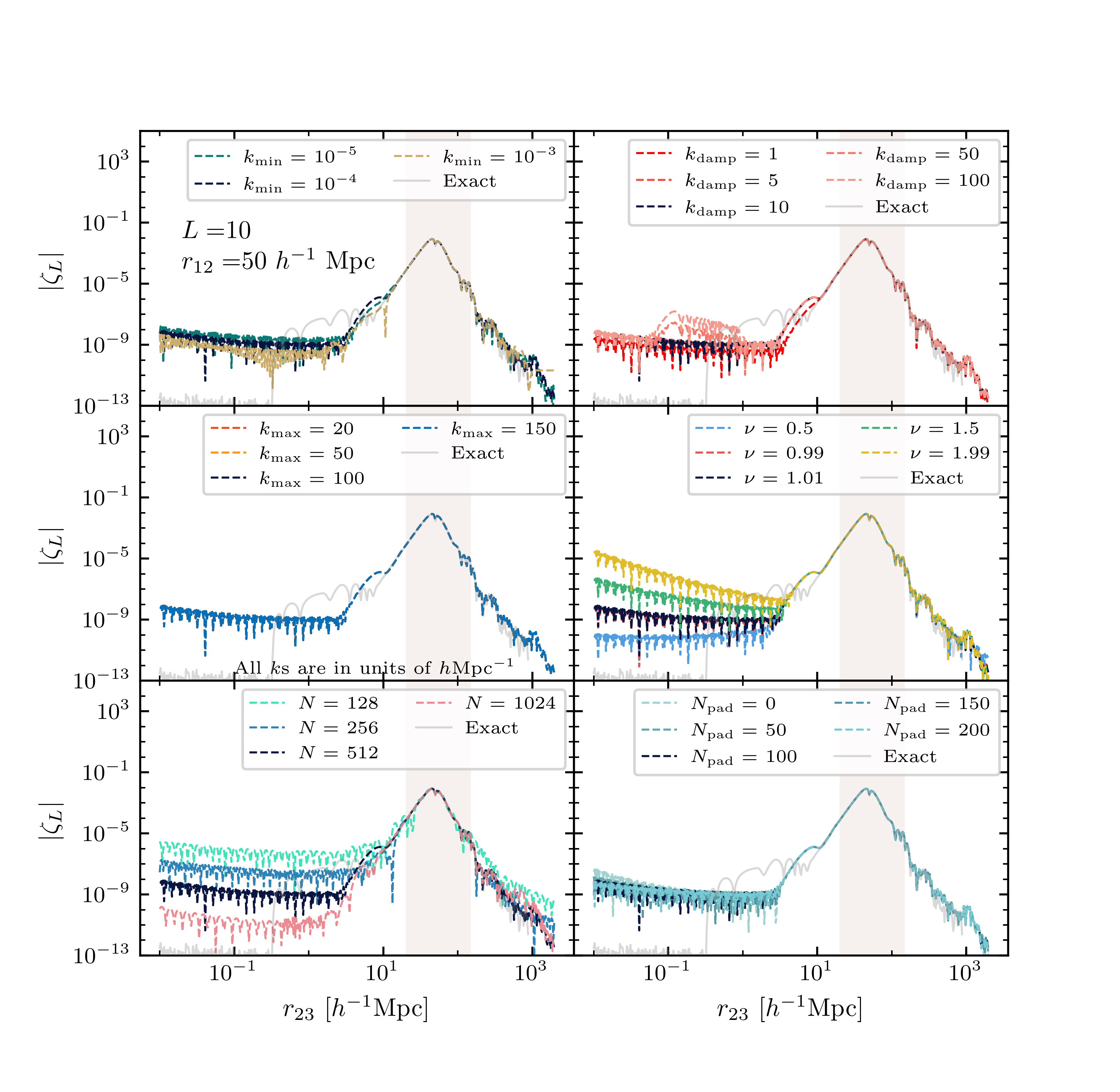}
    \caption{Impact of the parameters used in the 2D-FFTLog analysis, in particular the minimum scale $k_\mathrm{min}$ (top left), the maximum scale $k_\mathrm{max}$ (middle left), the length of the $k-$ and $r-$arrays $N$ (bottom left), the damping scale $k_\mathrm{damp}$ (top right), the ``bias'' parameter $\nu$ (middle right), and the length of the zero-padding array on each side of the $k-$array, $N_\mathrm{pad}$ (bottom right).
    All panels show the multipole $L=10$ of the real-space 3PCF for configurations with fixed $r_{12}= 50 \, h^{-1}\,\mathrm{Mpc}$.  In each panel, the dark blue line corresponds to the fiducial parameter choices, while all other lines vary a single parameter at a time.  The gray line is the reference computation using the real-space 3PCF from Appendix~\ref{sec:app.real-space-tree}.  The vertical band shows the range of scales we are interested in for the 3PCF comparison of Section \ref{sec: results}, i.e. $20-150 \, h^{-1}\,\mathrm{Mpc}$.
 }
    \label{Fig: 2DFFTLog}
\end{figure}

We find that the choice of the 2D-FFTLog parameters affects the higher-order multipoles more strongly, which is why we show the results in Figure~\ref{Fig: 2DFFTLog} for the highest multipole considered in this work, $L=10$.  We fix the scale $r_{12} = 50\,h^{-1}\,\mathrm{Mpc}$ and vary $r_{23}$, where the range of values relevant to our analysis in the main text is indicated by the vertical band.  In each panel, the dark blue dashed line represents our fiducial parameter choices.

The left column of Figure \ref{Fig: 2DFFTLog} shows the impact of changing the minimum and maximum $k$-scales used for the integration, as well as the number of bins $N$ (or equivalently, the logarithmic spacing).  The impact of $k_{\mathrm{max}}$ is irrelevant, and we fix it to $k_{\mathrm{max}}=100 \, h \, \mathrm{Mpc}^{-1}$. The impact of the minimum scale is distinguishable at very small and large scales, similarly for the dependence on the array length. For the main calculations we choose $k_{\mathrm{min}}= 10^{-4}\, h \,\mathrm{Mpc}^{-1} $ and $N=512$. In both cases the parameters are selected for the model to agree with the exact prediction up to sufficiently small scales, not to potentially impact the analysis, and at the same time to avoid excessive increase of computational times.

In order to avoid ringing and aliasing effects, caused by abrupt truncation of the function to Fourier-transform, and the inadequate accounting for the periodicity of the matrix assumed in the FFT algorithm, we pad the matrix adding $N_\mathrm{pad}$ null values at each side and each dimension of the matrix, and we damp the power spectrum at a scale $k_\mathrm{damp}$ with a Gaussian damping function, i.e. $P(k) = P_\mathrm{lin}(k) \exp[- (k/k_\mathrm{damp})^2]$. The damping of the power spectrum is also applied to the 1D integrals of the exact computation, in order to avoid spurious oscillatory behaviours, with a damping scale of $k_\mathrm{damp}= 10 \, h \, \mathrm{Mpc}^{-1}$.

The impact of the damping scale is shown in the upper right panel of Figure \ref{Fig: 2DFFTLog}. In the extreme case with $k_\mathrm{damp}=1 \, h \mathrm{Mpc}^{-1}$, the 3PCF multipole is damped at scales of almost order $10 \, h^{-1}$ Mpc. On the opposite, if the damping scale is close to the maximum scale of the computation, i.e. for $k_\mathrm{damp}=50 \, h \, \mathrm{Mpc}^{-1}$ and $k_\mathrm{damp}=100 \, h \, \mathrm{Mpc}^{-1}$, the damping has no impact and we see a spurious maximum at small scales. 
The value chosen for the main analysis is $k_\mathrm{damp}=10 \, h \, \mathrm{Mpc}^{-1}$.

The ``bias'' parameters $\nu_1$ and $\nu_2$ have been taken to be identical $\nu_1=\nu_2=\nu$, and tested in the allowed range $-L < \nu < 2$. The effect is particularly visible at small scales, where for larger values of the bias we see a deviation from the exact prediction at larger scales. 
At configurations with smaller fixed triangle side $r_{12}$, a deviation at large scales is also present for small bias values.  In the main analysis the value chosen is the one also adopted in \cite{fang2021}: $\nu = 1.01$. 

The number of null $k-$points added in the zero-padding has also been tested, as shown in the bottom right panel of Figure \ref{Fig: 2DFFTLog}. A small impact is visible, specifically the difference between the case with no zero-padding $N_\mathrm{pad}=0$ and the other cases is distinguishable. The parameter chosen in the main analysis is $N_{\mathrm{pad}}=100$. 

Note that, in all panels, when the 2D-FFTLog results differ from the exact prediction, the model shows strongly oscillatory behaviours. These deviations, however, are in all cases well outside the interval of scales we are interested in in the main analysis: there the impact of the parameters is not relevant for our purpose. We are therefore confident that the analysis will be consistent for slightly different values of the 2D-FFTLog parameters.

\end{document}